\documentstyle[aps,epsfig,eqsecnum,amsfonts]{revtex}

\draft

\begin{document}

\title{Numerical simulation of electromagnetic wave scattering from
    planar dielectric films deposited on rough perfectly conducting substrates}

\author{Ingve Simonsen\footnote{Email : Ingve.Simonsen@phys.ntnu.no}}
\address{Department of Physics, Theoretical Physics Group \\
The Norwegian University of Science and Technology (NTNU), \\
N-7034 Trondheim, \\
Norway}

\author{Alexei A. Maradudin\footnote{Email : aamaradu@uci.edu}}
\address{Department of Physics and Astronomy, and Institute for
    Surface and Interface Science,\\
    University of California, Irvine, CA 92697,\\
    USA}

\date{\today}
\maketitle


\begin{abstract}
    Electromagnetic wave scattering from planar dielectric films deposited
    on one-dimensional, randomly rough,
    perfectly conducting substrates is  studied by numerical
    simulations for both p- and s-polarization.
    The reduced Rayleigh equation, which is the integral equation
    satisfied by the scattering amplitude after eliminating the fields
    inside the film, is the starting point for the simulation.
    This equation is solved numerically by considering a random
    surface of finite length, and by introducing wave number
    cut-offs in the evanescent part of the spectrum.
    Upon discretization, a system of linear equations is obtained, and
    by solving this matrix system for an ensemble of surface
    realizations, the contribution to the
    mean differential reflection coefficient from the incoherently
    scattered field,
    $\left<\partial R_\nu/\partial\theta \right>_{\mathrm incoh}$ 
    $(\nu\!=\!p,s)$, is obtained nonperturbatively.
    It is  demonstrated that    when the scattering geometry supports
    at least two guided waves, 
    $\left< \partial R_\nu/\partial\theta\right>_{\mathrm incoh}$,
    has, in addition to the well known enhanced backscattering peak,
    well-defined satellite peaks in agreement  with theory, 
    for most of the parameters considered.
\end{abstract}


\section{Introduction}

Over the last five years, the
scattering of light from bounded systems where one of the interfaces
is rough has been studied in several papers. It was demonstrated that these
systems may give rise to
special enhancements in the angular distribution of the
 intensity of the light scattered incoherently, in addition to the
more well-known enhanced backscattering peak~\cite{Jun_Lu,PhysRep}
which is known to appear in the retroreflection direction.
These enhancements, known as satellite peaks,  are present only for bounded
systems supporting two (and not many more)
surface or guided waves at the frequency $\omega$
of the incident light~\cite{Freilikher94A}.
They are the result of the coherent interference of multiply-scattered
waves that are time-reversed partners of each other.

The scattering systems previously considered
in the literature in connection with the study of satellite peaks
naturally divide into two classes. The first one consists
of a rough dielectric film deposited on a planar perfectly conducting
substrate~\cite{Freilikher94A,Sanchez-Gil94,Wang,Sanchez-Gil96}.
The second geometry is a thin metal plate whose
illuminated (top) surface is a randomly rough one-dimensional surface,
while the lower one is planar~\cite{Sanchez-Gil95,McGurn89}
(see also Ref.~\cite{Ogura}).
In addition, systems containing volume disorder, in
contrast to surface disorder,
have also been considered, but such systems will
not be treated in the present work~\cite{Freilikher94B}.
The formalisms applied to the study  of these scattering systems range from
perturbation theories, such as small-amplitude
perturbation theory~\cite{Sanchez-Gil94}
and many-body perturbation theory~\cite{Sanchez-Gil96,Sanchez-Gil95},
stochastic functional methods~\cite{Wang,Ogura},
and numerical simulations~\cite{Madrazo}.
It should be noted that when low-order perturbation theory is not
valid, a non-perturbative approach has to be taken. Such an approach
is provided by numerical simulation techniques like the one presented
in the present work.

The scattering geometries considered in the works  mentioned
above ~(except~Ref.\cite{Ogura})  have one thing in common,
namely that the illuminated surface is always  the randomly rough
surface, while the back surface is planar. In the
present paper  we consider a geometry where the lower interface is
randomly rough and the upper (illuminated) one is planar.
In particular, we consider
a planar dielectric film deposited on a perfectly conducting rough
substrate (see Fig.~\ref{Fig:1}).
Proceeding from the reduced Rayleigh equation for the scattering amplitude
we use
numerical simulations to obtain the contribution to the mean  differential
reflection coefficient from the incoherent component of the scattered light
(also known as the mean incoherent differential
scattering cross section).

This paper is organized as follows.  We start by describing the
scattering geometry, the statistical properties of the rough
surface, and the satellite peaks supported by this geometry
(Sec.~\ref{sec:scattering geometry}).
In Sec.~\ref{Sec:RRE}, the reduced Rayleigh equation is derived,
and in Sec.~\ref{Sec:numerical method} the numerical method used in
the present work for its solution is discussed.
The numerical results and their discussion are presented in
Sec.~\ref{Sec:Results and discussions}.
Finally, in Sec.~\ref{Sec:Conclusions} the conclusions from these results
are presented.

\section{The scattering geometry}
\label{sec:scattering geometry}

The scattering geometry we consider in this work consists of a
dielectric film, characterized by the dielectric function
$\varepsilon(\omega)=\varepsilon_1(\omega)+i\varepsilon_2(\omega)$,
deposited on  a  randomly rough perfectly conducting semi-infinite
substrate. The top interface of the dielectric film is planar,
and the medium above the film is assumed for simplicity to be vacuum.
The mean thickness of the film is $d$. The scattering system is depicted in
Fig.~\ref{Fig:1}. The interface separating the perfect conductor and the
film is described by the surface profile function $\zeta(x_1)$.
This function is assumed to be a single-valued function of $x_1$.
Furthermore, it will be assumed to be a zero-mean, stationary, Gaussian
random
process defined through the following properties
\begin{eqnarray}
    \label{profile-function}
    \left< \zeta(x_1) \right> &=& 0,     \\
    \left< \zeta(x_1) \zeta(x'_1) \right>
            &=& \sigma^2 W(\left|x_1-x'_1 \right|).
\end{eqnarray}
Here $\sigma$ is the {\sc rms}-height of the surface roughness, and
$W(|x_1|)$ is the surface-height autocorrelation function
normalized such that $W(0) = 1$.
The angle brackets denote an average over the ensemble of realizations of
the surface profile
function $\zeta(x_1)$.
It will later on prove useful to have the power spectrum,  $g(|k|)$,
of the surface roughness at our disposal. It is defined as
\begin{eqnarray}
    \label{psd}
    g(|k|)  &=& \int^{\infty}_{-\infty} dx_1  W(|x_1|)\;  e^{-ikx_1}.
\end{eqnarray}
In the present  paper two forms of  the power spectrum will be considered.
These
are the Gaussian power spectrum defined by
\begin{eqnarray}
    \label{psd-gaussian}
        g(|k|)  &=& \sqrt{\pi}a \; e^{-\frac{k^2 a^2}{4}},
\end{eqnarray}
where $a$ is the transverse  correlation length of the surface roughness,
and the West-O'Donnell power spectrum
\begin{eqnarray}
    \label{psd-west-o'Donnell}
    g(|k|)  &=& \frac{\pi}{k_{+}-k_{-}}
    \left[  \theta(k-k_{-}) \theta(k_{+}-k)
          + \theta(-k_{-}-k) \theta(k+k_{+})\right] ,
\end{eqnarray}
where $\theta (k)$ is the Heaviside unit step function.  The latter power
spectrum was recently used by West and O'Donnell~\cite{West} 
in an experimental study of enhanced backscattering 
by the surface plasmon polariton mechanism in the scattering of 
p-polarized light from a one-dimensional, randomly
rough, metal surface.  In their work West and O'Donnell defined $k_+$ and
$k_-$ by
\begin{eqnarray}
k_{\pm} = k_{sp}(\omega ) \pm (\omega /c)\sin\theta_{\mathrm{wo}} ,
\end{eqnarray}
where $\pm k_{sp}(\omega )$ are the wave numbers of the forward- and
backward-propagating surface plasmon polaritons at a planar vacuum-metal
interface, whose frequency $\omega$ is that of the incident field.  The
physical significance of the angle $\theta_{\mathrm{wo}}$ is that if the angle of
incidence $\theta_0$ is in the interval 
$(-\theta_{\mathrm{wo}},\theta_{\mathrm{wo}})$, the
incident light can excite both the forward and backward propagating surface
plasmon polaritons through the surface roughness.  Similarly, if the
scattering angle $\theta$ is in the interval 
$(-\theta_{\mathrm{wo}},\theta_{\mathrm{wo}})$,
the excited surface plasmon polaritons will be coupled to scattered volume
waves in the vacuum region above the metal surface.  With this form of the
power spectrum the incident light couples strongly to the surface plasmon
polaritons over a limited range of angles of incidence rather than weakly
over a large range of this angle, as is the case when a power spectrum
$g(|k|)$ peaked at $k = 0$, e.g. a Gaussian, is used.  This is because in
the latter case the wave numbers $\pm k_{sp}(\omega )$ lie in the wings of
the power spectrum, where it is usually small.  As a consequence of the
strong excitation of surface plasmon polaritons by the incident light when
the power spectrum~(\ref{psd-west-o'Donnell}) is used, 
the amplitude of the enhanced peak, which is caused by the 
coherent interference of multiply scattered surface
plasmon polaritons with their time-reversed partners, is significantly
increased with respect to its value when the Gaussian power 
spectrum~(\ref{psd-gaussian}) is used.

To get an intuitive picture of these surfaces it is  useful to
supply the mean slope, $s$, and the mean distance
between consecutive peaks and valleys, $\left<D\right>$, as measured
along the (lateral) $x_1$-direction.
For a stationary zero-mean, Gaussian random process,
the {\sc rms}-slope $s$ is related to the power spectrum by
\begin{eqnarray}
    \label{rms-slope}
    s &=& \left<(\zeta'(x_1))^2\right>^{1/2}
       =  \sigma \sqrt{\int_{-\infty}^{\infty} 
               \frac{dk}{2\pi} \; k^2  g(|k|)},
\end{eqnarray}
and a good estimator for $\left<D\right>$ has been shown to be~\cite{Michel90}
\begin{eqnarray}
    \label{peaks-valleys}
    \left<D\right> &\simeq&
    \pi
    \sqrt{
          \frac{\int_{-\infty}^{\infty} dk \; k^2 g(|k|)}{
                 \int_{-\infty}^{\infty} dk\; k^4 g(|k|)}
         }.
\end{eqnarray}
For the two power spectra considered in this work, these quantities are
given by
\begin{eqnarray}
      s   &=& \left\{
          \begin{array}{ll}
              \sqrt{2}\frac{\sigma }{a} & \qquad \mbox{Gaussian} \\
              \;\frac{\sigma }{\sqrt{3}}\sqrt{k_+^2+k_+k_-+k_-^2}
              & \qquad \mbox{West-O'Donnell}
          \end{array}
        \right. ,
\end{eqnarray}
and
\begin{eqnarray}
    \left<D\right> &=&  \left\{
          \begin{array}{ll}
              \frac{\pi}{\sqrt{6}}a & \qquad \mbox{Gaussian} \\
              \pi \sqrt{ \frac{5}{3}
                          \frac{k_+^3-k_-^3}{k_+^5-k_-^5}}
              & \qquad \mbox{West-O'Donnell}
          \end{array}
        \right. .
\end{eqnarray}
Two surface profiles with the same (Gaussian) height
distribution, but with a Gaussian and a West-O'Donnell power spectrum
possessing nearly the same value of the {\sc rms}-slope $s$
are plotted in Fig.~\ref{Fig:2}.

\subsection{Satellite peaks}
\label{Sec:Satellite peaks}

Satellite peaks appearing in the scattering of light from bounded systems
are well-defined enhancements in the angular distribution of the
scattered (or transmitted) intensity
evenly distributed around the enhanced backscattering peak.
They are multiple-scattering phenomena, and are a consequence
of the constructive interference of multiply-scattered waves with their
time-reversed partners. In order for   satellite peaks to exist, the
scattering system must
support $N$ guided waves, where $N$ is at least two.
If we denote the  wavenumbers of these $N$
guided waves by $q_1(\omega)$,$q_2(\omega)$, $\ldots$, $q_N(\omega)$,
it can be demonstrated that in the absence of
roughness and absorption ($\varepsilon_2(\omega)=0$),
the satellite peaks should appear at scattering angles defined by
the following relation~\cite{Sanchez-Gil94,Sanchez-Gil96}
\begin{eqnarray}
    \label{satellite-peaks}
    \sin \theta_{(m,n)}^\pm
           &=& -\sin\theta_0 \pm
                \frac{c}{\omega}\left[q_m(\omega)-q_n(\omega)\right],
            \qquad  m\neq n.
\end{eqnarray}
It should be stressed that not all of these angles may correspond to
real satellite peaks. Some of them may lie in the evanescent
(non-radiative) part of the
spectrum, i.e. the right hand side of Eq.~(\ref{satellite-peaks}) 
may be greater than unity in magnitude. 
Furthermore,   not all of the real satellite peaks are
guaranteed to be observed, due to their low intensity.

In the present work when the power spectrum~(\ref{psd-west-o'Donnell}) 
is assumed, we will choose the values of 
$k_+$ and $k_-$ to include the wavenumbers of selected
guided waves supported by the scattering system depicted in Fig. 1,
enhancing the roughness-induced excitation of these guided waves as a
result.  We will see that the amplitudes of the corresponding satellite
peaks, as well as the amplitude of the enhanced backscattering peak, will
be increased thereby, in comparison with their values when a Gaussian power
spectrum is used.

It should be obvious that the wavenumbers of the guided waves,
$q_n(\omega)$, are related to the mean thickness, $d$, of the
dielectric film. By applying the wave equation with the proper
boundary conditions, the dispersion relation for the guided waves can be
obtained. In the case of no roughness, i.e. $\zeta(x_1)=0$,
and no absorption, it is given by
\begin{eqnarray}
    \label{dispersion-p}
    \varepsilon(\omega) \beta_0(q,\omega)
           &=& \alpha(q,\omega)\tan\alpha(q,\omega)d
\end{eqnarray}
for p-polarization~\cite{Sanchez-Gil96}, and by
\begin{eqnarray}
    \label{dispersion-s}
    \beta_0(q,\omega) &=& \alpha(q,\omega)\cot\alpha(q,\omega)d
\end{eqnarray}
for s-polarization~\cite{Sanchez-Gil94},  with
$\beta_0(q,\omega)=\sqrt{q^2-(\omega^2/c^2)}$ and 
\begin{eqnarray}
  \label{alpha} 
  \alpha(q,\omega) &=& \sqrt{\epsilon (\omega )\frac{\omega^2}{c^2}-q^2},
        \qquad Re\,\alpha(q,\omega )>0,\; Im\,\alpha (q,\omega ) > 0.
\end{eqnarray}
From these dispersion relations, the number of guided waves supported
by the scattering geometry can be deduced.  This number $n$ is
controlled  by the following inequalities for
p-polarization~\cite{Sanchez-Gil96}
\begin{eqnarray}
    \label{num-sp-p}
    \frac{n-1}{2\sqrt{\varepsilon-1}}  < \frac{d}{\lambda} <
    \frac{n}{2\sqrt{\varepsilon-1}},
\end{eqnarray}
and for s-polarization~\cite{Sanchez-Gil94}
\begin{eqnarray}
    \label{num-sp-s}
    \frac{2n-1}{4\sqrt{\varepsilon-1}}  < \frac{d}{\lambda} <
    \frac{2n+1}{4\sqrt{\varepsilon-1}},
\end{eqnarray}
where $n=1,2,3,\ldots$ is the number of guided
modes supported by the system at the wavelength $\lambda$ of the
incident light.

\section{The reduced Rayleigh equation}
\label{Sec:RRE}

In the present work we will consider a plane wave incident from the
vacuum side on the structure described earlier~(Fig.~\ref{Fig:1}), with the
$x_1x_3$-plane  the plane of incidence.
By this choice for the  plane of incidence, we are guaranteed
that there is no cross-polarized scattering, and the plane of scattering
coincides with the plane of incidence.  As a consequence,
 {\em one} field component is sufficient to  describe the
electromagnetic field completely. Thus, to simplify the notation we introduce
\begin{eqnarray}
    \label{fields}
    {\mathbf \Phi}_\nu({\mathbf x},t)
      &=& \left\{
               \begin{array}{ll}
                  { \mathbf H}({\mathbf x},t)
                  & \quad \nu=p, \\
                  {\mathbf E}({\mathbf x},t)
                  & \quad \nu=s,\\
               \end{array}
           \right. \\
      &=&  \Phi_\nu(x_1,x_3 | \omega )\; e^{-i\omega t}
            \hat{\mathbf e}_2,
\end{eqnarray}
where $\hat{\mathbf e}_2$ is a unit vector in the $x_2$-direction, and
the subscript $\nu$ is a polarization index ($\nu=p,s$).
We have explicitly  assumed a time-harmonic dependence of the fields here.

In the region above the film, $x_3>d$, the field consists of an incident and
scattered waves. It can be represented by
\begin{eqnarray}
    \label{rex}
    \Phi_\nu(x_1,x_3|\omega)  &=&
        e^{ikx_1-i\alpha_0(k,\omega)x_3}
      + \int^{\infty}_{-\infty}
           \frac{dq}{2\pi} R_\nu(q|k) e^{iqx_1+i\alpha_0(q,\omega)x_3},
\end{eqnarray}
where $R_\nu(q|k)$ is the scattering amplitude, and
\begin{eqnarray}
    \label{alpha0}
    \alpha_0(q,\omega) &=& \left\{
      \begin{array}{ll}
          \sqrt{\frac{\omega^2}{c^2}-q^2}, & \quad |q|<\frac{\omega}{c},
          \\
          i \sqrt{q^2-\frac{\omega^2}{c^2}}, & \quad |q|>\frac{\omega}{c}.
      \end{array}
      \right.
\end{eqnarray}
Furthermore, the field inside the film consists of both  upward and downward
propagating modes. By applying the  boundary conditions
satisfied by the field at the
vacuum-dielectric and the dielectric-metal interfaces, two coupled
integral equations are obtained (the Rayleigh equations).
In obtaining these equations the Rayleigh hypothesis has been
imposed~\cite{Rayleigh96,Rayleigh07}.
The condition for the  validity of this hypothesis can crudely be stated
as~\cite{Ogilvy} $|\zeta'(x_1)|\ll 1$.
Now, by eliminating the modes inside the film,
a single integral equation for  $R_\nu(q|k)$, known as the
reduced Rayleigh equation, is obtained.
For the scattering system considered in the present work
it assumes  the following form~\cite{Simonsen98}
 \begin{eqnarray}
    \label{rre}
    \int^{\infty}_{-\infty}
        \frac{dq}{2\pi} M_{\nu}^{+}(p|q) R_\nu(q|k)
     &=& M^{-}_\nu(p|k),
\end{eqnarray}
where
\begin{eqnarray}
    \label{rre-matrix-elements-p-pol}
    M_{p}^{\pm}(q|p)
      &=& \pm e^{\pm i\alpha_0(q,\omega)d}\;
           \frac{\varepsilon\frac{\omega^2}{c^2}-pq}{\alpha(q,\omega)}
           \left[ e^{-i\alpha(q,\omega)d} \;
              \frac{\alpha(q,\omega)\pm\varepsilon(\omega)\alpha_0(q,\omega)
                   }{\alpha(q,\omega))}
                    I\left(\alpha(q,\omega)|p-q\right)
      \right. \nonumber \\  && \qquad + \left.
                  e^{i\alpha(q,\omega)d} \;
                 \frac{\alpha(q,\omega)\mp\varepsilon(\omega)\alpha_0(q,\omega)
                     }{-\alpha(q,\omega))}
                  I\left( -\alpha(q,\omega)|p-q\right) \right]
\end{eqnarray}
for p-polarization, and
\begin{eqnarray}
    \label{rre-matrix-elements-s-pol}
    M_{s}^{\pm}(q|p)
      &=& \pm e^{\pm i\alpha_0(q,\omega)d} \;
           \left[ e^{-i\alpha(q,\omega)d}\;
              \frac{\alpha(q,\omega)\pm\alpha_0(q,\omega)
                   }{\alpha(q,\omega))}
                    I\left(\alpha(q,\omega)|p-q\right)
      \right. \nonumber \\ &&  \qquad - \left.
                  e^{i\alpha(q,\omega)d} \;
                 \frac{\alpha(q,\omega)\mp\alpha_0(q,\omega)
                     }{-\alpha(q,\omega))}
                  I\left( -\alpha(q,\omega)|p-q\right) \right]
\end{eqnarray}
for s-polarization.
The function $I(\gamma |q)$ appearing in these  formulae is defined by
\begin{eqnarray}
    \label{int}
    I(\gamma|q)
         &=&  \int^{\infty}_{-\infty} dx_1
              e^{i\gamma\zeta(x_1)} e^{-iqx_1}, 
\end{eqnarray}
and $\alpha(q,\omega)$ was defined earlier in Eq.~(\ref{alpha}).
The main purpose of this paper is to calculate the scattering
amplitude $R_\nu(q|k)$ numerically.
This amplitude is related to a physically
measurable quantity, the mean differential reflection coefficient
$\left< \partial R_\nu /\partial\theta \right>$, which is defined as
the fraction of the total incident flux that is scattered into a small
angular interval $d\theta$ around the scattering direction $\theta$.
Since this quantity for
weakly rough surfaces will have a dominating peak due to coherent (specular)
reflection, it is useful to subtract off this contribution.
If we do so, we are left with what is called the contribution to the
mean differential reflection coefficient from the incoherent component of
the scattered light, which we will
denote by $\left< \partial R_\nu/\partial\theta \right>_{\mathrm incoh}$.
This quantity is related to the scattering amplitude $R_{\nu}(q|k)$ according
to~\cite{PhysRep}
\begin{eqnarray}
    \label{eq:mdrc-incho}
    \left< \frac{\partial R_\nu}{\partial \theta } \right>_{\mathrm incoh}
    &=& \frac{1}{L}\; \frac{\omega }{2\pi c}\;
        \frac{\cos^2{\theta}}{\cos{\theta_0}}\;
        \left[  \langle\left| R_\nu \right|^2 \rangle
               - \left| \left< R_\nu \right> \right|^2 \right].
\end{eqnarray}
Here $L$ is the length of the sample along the
$x_1$-direction, and $\theta_0$ and $\theta$ are the angles of incidence and
scattering, respectively.
These angles are defined in the counter and clockwise
directions as indicated in Fig.~\ref{Fig:1}, and they are
related to the wave numbers $k$ and $q$ by
\begin{eqnarray}
    \label{momenta1}
    k  = \frac{\omega}{c}\sin\theta_0, \qquad
    q  =  \frac{\omega}{c}\sin\theta .
\end{eqnarray}

\section{The numerical method}
\label{Sec:numerical method}

In the paper by Madrazo and Maradudin~\cite{Madrazo},
 their  reduced Rayleigh equation was solved numerically by replicating
the rough surface of length $L$ an infinite number of times.
By doing so, they covered the entire $x_1$-axis and obtained a
diffraction grating of period $L$. Hence the wave number  integration in
their equivalent of Eq.~(\ref{rre}) was converted into an infinite,
but for  practical implementation a  large but finite, sum over Bragg
beams. By this method they were able to obtain convergent results for
$\langle \partial R_{\nu}/\partial\theta\rangle_{\mathrm incoh}$.

In the present paper, we have chosen probably the most straightforward
method possible for the numerical solution of the reduced Rayleigh equation.
We do not replicate the surface periodically, but instead truncate
the wave number integration in Eq.~(\ref{rre}) somewhere in the evanescent
(non-radiative)
part of the spectrum, say at $q = \pm Q/2$, where $Q \gg \omega /c$.
With a finite length of the surface $L\gg\lambda$, the
reduced Rayleigh equation can now be discretized
by a standard quadrature scheme, and  a system of linear
equations is obtained. The result is
\begin{eqnarray}
    \label{discretized-rre}
    \frac{h_q}{2\pi} \sum^{N_q/2}_{n=-N_q/2}
        w_n M_{\nu}^{+}(p_m|q_n) R_\nu(q_n|k)
     &=& M^{-}_\nu(p_m|k), \qquad m=-\frac{N_q}{2},\ldots,\frac{N_q}{2},
\end{eqnarray}
where $N_q+1$ is the number of discretization points in wave number  space,
$h_q=Q/N_q$ is the corresponding discretization length,
and  $\{w_n\}$ are the weights of the quadrature scheme used.
Furthermore, the abscissas $\{q_n\}$ are defined
by
\begin{eqnarray}
    \label{mom-discretized}
    q_n &=& n h_q, \qquad n=-\frac{N_q}{2},\ldots,\frac{N_q}{2}.
\end{eqnarray}
The quadrature scheme used in the numerical results presented in the
next section is an ${\cal O}(h^4_q)$ method with~\cite{Press} $w_1=w_N=3/8$,
$w_2=w_{N-1}=7/6$, $w_3=w_{N-2}=23/24$, and with all other weights equal
to one. The matrix elements $M_{\nu}^{\pm}(p_m|q_n)$
in Eq.~(\ref{discretized-rre}), are given by
Eqs.~(\ref{rre-matrix-elements-p-pol}) and (\ref{rre-matrix-elements-s-pol}),
but now with a finite length of the surface, viz.
\begin{eqnarray}
    \label{int-mod}
  I(\gamma|q)
         &=&  \int^{\frac{L}{2}}_{-\frac{L}{2}} dx_1
              e^{i\gamma\zeta(x_1)} e^{-iq x_1}.
\end{eqnarray}
Integrals of this form are often referred to as Fourier integrals.
With finite wave number cut-offs at $q = \pm Q/2$,
it follows from Eqs.~(\ref{rre})--(\ref{rre-matrix-elements-s-pol})
that  wave numbers in the range $[-Q,Q]$ are needed in the
calculation of these integrals.
Since, as we will see in the following subsection,
these integrals will be evaluated by discretizing the spatial
$x_1$-variable and taking advantage of the (discrete) Fourier transform,
the value of $Q$ is controlled
by the spatial discretization length $h$ of the problem.
To resolve wave numbers up to $Q$, one has from the relation for critical
sampling (Nyquist frequency) that the number of spatial discretization
points has to satisfy $\pi/h\geq Q = h_qN_q$, where we have
used Eq.~(\ref{mom-discretized}). In principle the wave number
discretization length used in Eq.~(\ref{discretized-rre}), $h_q$, and
the one obtained from the Fourier transform, $h_q'$, are
independent.  However, from a numerical point of view this is not very
practical, and we will chose them to be the same, $h'_q =
h_q$. From the theory of the Fourier transform  we know that the
discretization lengths in real and wave number space are related by
$h_q=2\pi/(N h)$, where $N$ is the number of spatial discretization
points. If this latter expression is used in $\pi/h\geq Q = h_qN_q$,
one arrives at
\begin{eqnarray}
    \label{N-dim}
    N &\geq& 2N_q.
\end{eqnarray}
In the numerical calculation we have chosen $N=2N_q$ in order to
avoid unnecessary calculations.

\subsection{The integrals $I(\gamma|q)$}

In order to calculate the Fourier integrals in Eq.~(\ref{int-mod}), care
has to be taken. The reason for this is the oscillatory nature of these
integrals which appears unavoidable for large absolute values of the
wave number parameter $q$. For example, if one use a standard
discretization approach to this integral,
$I(\gamma|q_n)=h\cdot exp(iq_n L/2) \sum_{m=0}^{M-1}
{\cal F} [exp(i\gamma\zeta_{m})]$, where ${\cal F}$ denotes the
Fourier transform, one will likely obtain inaccurate and even wrong
results (see Ref.~\cite{Press} Sec. 13.9 for details).
In a more sophisticated and reliable method, the integrand is
approximated by higher-order (piecewise Lagrange)
interpolating  polynomials.
By taking advantage of the additional information
provided by these interpolating polynomials,
highly accurate results for the oscillating Fourier-integrals can be
obtained from
\begin{eqnarray}
    \label{endpoint-coorections}
    I(\gamma|q_n) &\simeq& h\; e^{iq_n\frac{L}{2}}
             \left[ W(q_nh) {\cal F} [\{e^{i\gamma\zeta_n}\}]
             + \sum_{j'}\Gamma_{j'}
             e^{i\gamma\zeta_{j'}}
            \right],
\end{eqnarray}
where the $j'$-summation extends over the endpoints of the integration
interval.
This endpoint correction appearing in Eq.~(\ref{endpoint-coorections})
enters due to the difference in the  functional form of the interpolating
polynomials for internal and boundary points.
The specific form of $W(\cdot)$ and $\Gamma_{j'}$ depends only on the
order of the interpolating polynomials. In this work cubic order
has been used, and the interested reader is referred to
Ref.~\cite{Press} for explicit expressions for these functions.

The most time consuming part of the numerical solution of the reduced
Rayleigh equation is not, as one might
expect at first glance, the solution of the matrix system. Instead
it is to set up this  system and, in particular, to
calculate the integrals $I(\gamma|q)$.
We stress that these integrals in principal have to be calculated
for each new value of $q$. This means, for the method just sketched,
that the Fourier transform
has to be recalculated for each new value of the wave number variable
$q$. Even though the Fourier transform has an fast implementation, the
fast Fourier transform (FFT), there are so many transforms to be
calculated that the whole operation becomes quite time consuming.
For a system of linear size $N$, the number of integral evaluations needed
is $2 N^2$, where  the factor of 2 enters because one needs two integral
evaluations per matrix element.
Another method, which overcomes this drawback when applicable, was
introduced in Ref.~\cite{Madrazo}, and consists of expanding the
exponential function, $exp(i\gamma\zeta(x_1))$ appearing in
Eq.~(\ref{int-mod}) in a Taylor series, and
integrating the resulting expression term-by-term
\begin{eqnarray}
    \label{Taylor}
    I(\gamma|q)
    &=&
        \int^{\frac{L}{2}}_{-\frac{L}{2}} dx_1 \;
        e^{-iq x_1} \sum_{n=0}^{\infty} \frac{(i\gamma)^n}{n!}
        \zeta^n(x_1) \nonumber \\
    &=&
        \sum_{n=0}^{\infty} \frac{(i\gamma)^n}{n!}
        {\cal F} [\zeta^n](q).
\end{eqnarray}
For surface profiles of modest roughness, this expansion will converge
rather quickly, meaning that only a limited number of terms are needed
 in the summation in Eq.~(\ref{Taylor}). However, the real
advantage of this method  is that the
Fourier-transforms needed to calculate $I(\gamma|q)$  do not have to
be recalculated for every new value of the wavenumber variable $\gamma$,
in contrast to what is the case for the Fourier-method presented above.
All this, as we will see, results in a dramatic reduction
of computational time (cf. Fig.~\ref{Fig:6}).

For the roughness used in the present work, the latter method for the
calculation of $I(\gamma|q)$ will be the method of choice. The
Fourier-method has been included and used to demonstrate and compare the
results obtained by its use to those obtained by the Taylor-method.
However, the  Fourier-method is the method of choice when the Taylor-method
begins to converge slowly, which can occur for large-amplitude roughness.
When we are forced to use the Fourier-method, high demands are placed on
computer time.

\section{Results and discussions}
\label{Sec:Results and discussions}

In this section we present the results of the simulations obtained by
the method just described.
For all simulations incident light of wavelength
$\lambda=633 $~nm was used. The rough surface had length
$L=160.1\lambda=101343.3$~nm, and an {\sc rsm}-height
$\sigma = 30$~nm, unless indicated otherwise.
The rough surfaces were generated by the method
described in Refs.~\cite{PhysRep,AnnPhys}. The number of surface
realizations used in the ensemble average was $N_{\zeta}=3000$, if nothing
is said to the contrary. Such a large
number of samples was used in order to reduce the noise level
which such numerical calculations are plagued with.
If nothing else is said, the integration method used
in the calculation of the integrals $I(\gamma|q)$ was the
Taylor-method where ten terms were retained.
This was enough, as we will see, to obtain convergent results.
The calculations about to be presented were performed on an
SGI/Cray Orion 2000 supercomputer, and the typical {\sc cpu}-time
for this number of samples was roughly
30 hours.
Furthermore, the number of spatial discretization points was set to
$N=1604$ (and hence $N_q=802$).
Since only a limited number of Fourier transforms was
needed per sample when using the Taylor-method, the advantage of the
FFT-algorithm is rather marginal as compared to the system size.

In the absence of absorption, i.e. when the dielectric function is
real, the following unitarity condition (conservation of energy),
coming from the unitarity of the scattering matrix,
should be satisfied
 \begin{eqnarray}
   \int^{\frac{\omega}{c}}_{-\frac{\omega}{c}} \frac{dq}{2\pi} \;
   \frac{\alpha_0(q,\omega)}{\sqrt{\alpha_0(k,\omega)\alpha_0(p,\omega)}}
        R^*(q|k) R(q|p) &=& 2\pi\; \delta(k-p).
\end{eqnarray}
Numerical simulations with negligible absorption showed that
this condition was satisfied within 0.5\% in the case that $k=p$.
With a small non-vanishing imaginary part of the dielectric function
($\varepsilon_2(\omega)=0.01$), this relation was satisfied within
4\%--16\%, depending on the roughness used.

\subsection{S-polarization}

We start the presentation of the numerical results by
considering s-polarized incident light.
The dielectric constant of the film is,
at the given wavelength, $\varepsilon(\omega)=2.6896+i0.01$,
and the mean film-thickness is set to $d=500$~nm.
This is the same set of parameters used for s-polarization
in Ref.~\cite{Madrazo}.
In the absence of roughness and absorption, this mean film thickness
predicts, according to Eq.~(\ref{dispersion-s}), two guided
wave modes with wave numbers  $q_1(\omega)=1.5466(\omega/c)$ and
$q_2(\omega)=1.12423 (\omega/c)$ (cf. Ref.~\cite{Sanchez-Gil94}).
The corresponding satellite peaks are then, according to
Eq.~(\ref{satellite-peaks}),  expected to appear at
$\theta = \pm 17.7^\circ$.

In Fig.~\ref{Fig:3} we present the results of the numerical simulation of the
scattering of s-polarized light normally incident on a surface for which
both the surface height distribution and the surface-height autocorrelation
function have the Gaussian form. The correlation length of the surface
roughness is $a=100$~nm. With these parameters the {\sc rms}-slope
and mean distance between consecutive peaks and valleys are
given by
$s=0.424$ and $\left<D\right>=128.3$~nm, respectively.
In the raw data for the mean incoherent reflection coefficient
(Fig.~\ref{Fig:3}a) the satellite peaks
are hard to see due to the noisy background, but  the
enhanced  backscattering peak at $\theta =0^{\circ}$
is easily located.
This conclusion is not affected by doubling the number
of Taylor-terms retained in the calculation~(result not
shown). However, we believe that the noise stems from the fact that we
are using a plane incident wave instead of a beam of finite  width.
In order to reduce the effect of the noise, we have applied a standard
five-point filter to the raw data.
This filter affects only eleven consecutive points, and it should be noted
that this smoothing procedure  is not as aggressive as the one
used by Madrazo and Maradudin~\cite{Madrazo}.
After applying the smoothing-filter to the raw data,
the satellite peaks appear more clearly~(Fig.~\ref{Fig:3}b).
Even though their amplitudes are quite small,
they seem to appear at the correct
angles as indicated by the dashed line in the figure.
Satellite peaks, like enhanced backscattering peaks, are
multiple scattering effects, and can consequently be masked by
single-scattering phenomena. This is probably the reason for the small
amplitudes of the satellite peaks in Fig.~\ref{Fig:3}b.

We now focus on the West-O'Donnell power spectrum.
From perturbation theory~\cite{PhysRep}, it is known that the lowest
order contribution, i.e. the single-scattering contribution, to the
mean incoherent differential reflection coefficient is proportional to the
power spectrum $g(|q-k|)$,  where $q$ and $k$ are the wave numbers of the
scattered and incident waves, respectively
(cf. Eq.~(\ref{momenta1})). This  implies that the
single-scattering contribution to
$\left<\partial R_\nu/\partial\theta\right>_{\mathrm incoh}$  is largest for
angles where $g(|q-k|)$ is large. As mentioned earlier, West and
O'Donnell constructed a power spectrum $g(|k|)$ which is non-vanishing
only in a limited range of $|k|$ defined by
the upper and lower limits $k_+$ and $k_-$, respectively.
Thus, for angles of incidence and
scattering satisfying $|q-k|<k_-$,
single scattering processes will not contribute to $\langle \partial
R_{\nu}/\partial\theta \rangle_{\mathrm incoh}$, and may  enhance
the roughness-induced coupling of electromagnetic waves to surface
plasmon polaritons if the values of $k_\pm$ are properly chosen.
In the scattering process we are about to consider, we have chosen
$k_-=0.82 (\omega/c)$ and $k_+=1.97 (\omega/c)$, and thus
$s=0.427$ and $\left<D\right>=201.1$~nm, so that the incident
wave, for small  angles of incidence,
can excite both guided waves supported by the geometry
through the surface roughness.
With these parameters, and  for normal incidence,
one finds from Eq.~(\ref{momenta1})
 that single-scattering effects
do not contribute to the mean incoherent differential reflection
coefficient for scattering  angles $|\theta| <  55.1^\circ$.
Since the satellite peaks, located  at $\theta=\pm17.7^\circ$
in this case, cannot be masked by
single-scattering processes, one   now expects more
pronounced satellite peaks then those obtained with the use of the Gaussian
correlation function presented in Fig. 3.
As can be seen from Fig.~\ref{Fig:4}a, this is indeed what we get.
The satellite peaks are well separated from the background in the raw
data, and no smoothing is needed.
The amplitude of the prominent enhanced backscattering
peak  seen in this figure is twice that of its
background. This is another indication that single scattering
processes have not contributed~\cite{Madrazo24}.
The abrupt increase in
$\left<\partial R_s/\partial\theta\right>_{\mathrm incoh}$  for
angles $|\theta|$ somewhat  above $50^\circ$ is due to
single-scattering  effects setting in.
We predicted above that this should  happen for angles $|\theta| \geq
55.1^\circ$,
a result that fits quite well with what can be read off from the
figure.
In Fig.~\ref{Fig:4}b we have shown the (smoothed) result of the
simulation obtained by Madrazo and Maradudin~\cite{Madrazo} for the
same set of parameters, but with the top interface being the rough
one instead of the lower one.
It is observed that the overall amplitude is
three times smaller than  the one presented in Fig.~\ref{Fig:4}a.
Furthermore, the intensity obtained for our geometry in the single
scattering regime ($|\theta| > 55.1^\circ$) has a broader distribution than
the one
obtained by Madrazo and Maradudin~\cite{Madrazo}.
The reason for this has to do with the increased reflection at the rough
surface present in our case.

In the previous section we discussed two methods for the calculation
of the  integrals $I(\gamma|q)$. We will now compare these two methods
with respect to numerical performance and accuracy. We will use the
West-O'Donnell power spectrum with the parameters used above, but with the
number of spatial discretization points  chosen to be
$N=2^{11}=2048$. This is done in order to take advantage of the fast
Fourier transform.
In Fig.~\ref{Fig:6} the results for the  mean incoherent
reflection coefficients, $\left<\partial R_s/\partial
  \theta\right>_{\mathrm incoh}$ are presented for the Fourier-
(Fig.~\ref{Fig:6}a) and Taylor- (Fig.~\ref{Fig:6}b) methods.
The number of samples used in obtaining these results was $N_\zeta=50$.
For the Taylor-method, the number of terms retained
in the Taylor expansion~(\ref{Taylor}) was ten.
The {\sc cpu}-time needed to obtain these results on a
SGI/Cray Orion~2000 supercomputer were $t_{\mathrm F}=63.3$~{\sc cpu}-hours
and $t_{\mathrm T}=1.3$~{\sc cpu}-hours for the Fourier- and
Taylor-methods, respectively.
In Fig.~\ref{Fig:6}c the discrepancy between the results of the
two methods are shown. The relative error is roughly 1\%, which is of
the order of the error due to the use of a finite number of
samples~\cite{AnnPhys}.
Thus we may conclude that for the  parameters used, the results are
equivalent, and a factor of 50 in performance is gained
by using the Taylor-method instead of the Fourier-method.
Consequently, the Taylor-method will be the method of choice
for slightly rough surfaces, and we will use this method in the
following numerical calculations.
However, we should stress that this result is not a general one,
but depends heavily on the roughness parameters used.
In particular, for sufficiently large {\sc rms}-heights, the Taylor-method
will converge slowly, and we are left with the numerically demanding
Fourier-method.

As the  angle of incidence departs  from the direction
of normal incidence ($\theta_0=0^\circ$), the amplitudes of the satellite
peaks are known to
decrease.
In Fig.~\ref{Fig:5} we show the results of a simulation for a
geometry with the same parameters as above, but now with  angle of incidence
of $\theta_0=5^\circ$. According to Eq. (\ref{satellite-peaks})
the satellite peaks should now appear at $\theta=12.6^\circ$
and $\theta=-26.1^\circ$.
It is seen from the figure that the amplitudes of these peaks, and the
enhanced backscattering peak (at $\theta=5^\circ$),  are reduced in
amplitude compared to the case of normal incidence,
but they are still easily distinguished from the
background and are located at the predicted positions.

\subsection{p-polarization}

We will now focus our attention on p-polarization of the incident
light. Satellite peaks are typically harder to observe in
p- than in s-polarization. This is related to  the reduced reflectivity
at the rough surface of p-polarized light  as compared
to that of  s-polarization for the same angle of incidence.
In fact, this was the reason why Madrazo and
Maradudin~\cite{Madrazo} used different parameters for the two
polarizations considered.
However, in this paper, the rough surface is perfectly reflecting, and
we thus use the same set of parameters for the
two polarizations. For completeness and comparison, at the end of this
section we have included results for the parameters used by
Madrazo and Maradudin~\cite{Madrazo}.

We start by presenting the numerical results obtained for the
parameters used in the preceding subsection.
From Eq.~(\ref{num-sp-p}) one finds that the scattering system,
now in the case of p-polarization,
supports three guided waves, which according to the dispersion
relation~(\ref{dispersion-p}) have the wave-numbers
$q_1(\omega)=1.6125 (\omega/c)$, $q_2(\omega)=1.3821 (\omega/c)$ and
$q_3(\omega)=1.0029 (\omega/c)$. Hence there are six possible
satellite peaks, and for normal incidence they are located at scattering
angles (cf. Eq.~(\ref{satellite-peaks}))
$\theta_{\left( 1,2\right)}^\pm=\pm 13.3^\circ$,
$\theta_{\left(2,3\right)}^\pm=\pm 22.3^\circ$,
and $\theta_{\left(1,3\right)}^\pm=\pm 37.6^\circ$.

In Fig.~\ref{Fig:7} the contribution $\langle \partial R_{p}/\partial\theta
\rangle_{\mathrm incoh}$ is presented for a rough surface described by a
Gaussian power spectrum with correlation length $a=100$~nm.
As for the case of s-polarization with the same set of parameters,
the enhanced backscattering peak is easily seen
in the raw data (Fig~\ref{Fig:7}a). However, there is now clear
evidence of the satellite peaks in these data.
After smoothing ~(Fig.~\ref{Fig:7}b),
the satellite peaks at $\theta_{(2,3)}^\pm$ are clearly visible,
while there is only a weak indication of  those at $\theta_{(1,2)}^\pm$.
However, there is no indication of the satellite peaks at
$\theta_{(1,3)}^\pm$, neither in the raw, nor in the smoothed data.

We saw earlier that, in going to a surface roughness described by the
West-O'Donnell power spectrum, the satellite peaks became much more
pronounced because they were not masked by single-scattering
processes. Figure~\ref{Fig:8} presents the results for the  scattering of
normally ($\theta_0=0^\circ$) incident p-polarized light
when the power spectrum is of the
West-O'Donnell type with parameters $k_-=0.82(\omega/c)$
and $k_+=1.97(\omega/c)$ (as above).
Here, as in the case of  s-polarization,
the satellite peaks at $\theta_{(2,3)}^\pm$ are well defined both in
the raw (Fig.~\ref{Fig:8}a) and smoothed data (Fig.~\ref{Fig:8}b).
In the raw data, the peaks at $\theta_{(1,2)}^\pm$ can also be seen,
even though they appear  more  clearly in the smoothed  data.
As in the Gaussian case, there is no sign of the satellite
peaks at  $\theta_{\left(1,3\right)}^\pm$
neither in the raw nor in the  smoothed data.
This may indicate that the coupling of the incident wave to the guided
mode with wavenumber $q_3(\omega)$ is quite weak.
We observe that the dominant satellite peaks
($\theta_{\left(2,3\right)}^\pm$)
have roughly  the same amplitude as those observed for s-polarization.
This has to do with the fact, as we claimed above,
that the randomly rough surface is perfectly reflecting,
and thus no distinction between p- or s-polarization when it
comes to reflectivity has to be made.
However, for the geometry
considered by Madrazo and Maradudin~\cite{Madrazo}, where the rough
surface was the vacuum-dielectric interface, the amplitude of the
satellite peaks for p-polarization were much lower then those obtained in
s-polarization, due to the decreased reflectivity for
the p-polarized light as compared to that for s-polarized light.

If the incident light has an angle of incidence  of, say, $\theta_0=5^\circ$,
the satellite peaks should
appear at $\theta_{\left(1,2\right)}^\pm=8.2^\circ,-18.5^\circ$,
$\theta_{\left(2,3\right)}^\pm=17.0^\circ,-27.8^\circ$,
and $\theta_{\left(1,3\right)}^\pm=31.5^\circ,-44.2^\circ$,  according to
Eq.~(\ref{satellite-peaks}).
The results  for this situation, still with the West-O'Donnell
power spectrum,  are given in Fig.~\ref{Fig:9}a and \ref{Fig:9}b
for the raw and smoothed data, respectively.
Also in this case the peaks at  $\theta_{(1,2)}^\pm$
and $\theta_{(2,3)}^\pm$ are detectable.
However, it is interesting to note that the peaks at
$\theta_{(1,3)}^\pm$, of which there was no sign  for an angle of incidence
$\theta_0=0^\circ$,   are weakly visible   in both the raw
and smoothed data. This we find rather surprising, since typically
the satellite peak amplitudes decrease quite rapidly with increasing angle of
incidence.

In the Madrazo and Maradudin paper~\cite{Madrazo}, they used another
dielectric function, $\varepsilon(\omega)$, and mean film thickness,
$d$,  for p-polarization  than they (and we) used in the simulations
for s-polarization.
In particular, they used $\varepsilon(\omega)=5.6644+0.01i$ and
$d=380$~nm.  All other parameters were the ones used earlier in this
paper.
For the purpose of comparison, we have included  simulation results
for these parameters.
In Fig.~\ref{Fig:10} the result for
$\left< \partial R_p/\partial \theta\right>_{\mathrm incoh}$  is
presented for a surface possessing a Gaussian power spectrum with a
correlation length $a = 100$~nm (and with  $\sigma=30$~nm as before).
This corresponds to an {\sc rms}-slope of $s=0.424$, and the mean
distance between consecutive peaks and valleys is
$\left<D\right>=128.3$~nm.
For such a surface, in the limit of no roughness and no absorption
in the film ($\varepsilon_2(\omega)=0$), the structure supports
three guided modes whose wave numbers are $q_1(\omega)=2.34 (\omega/c)$
$q_2(\omega)=2.04 (\omega/c)$ and $q_3(\omega)=1.32 (\omega/c)$. Hence,
there are six satellite peaks, which according to
Eq.~(\ref{satellite-peaks}), are located at
$\theta_{\left(1,2\right)} = \pm17.5$, $\theta_{\left(2,3\right)} = \pm46.1$,
while $\theta_{\left(1,3\right)}$ lie in the evanescent
(non-radiative) part of the spectrum and thus are not visible.
From Fig.~\ref{Fig:10} it is seen
that even the satellite peaks in the radiative part of the spectrum are
not possible to locate in the numerical results. This is in fact the same
conclusion drawn by Madrazo and Maradudin~\cite{Madrazo}
for their geometry. Thus the increased reflectivity we have at the
rough surface does not seem to affect the
detection of the possible satellite peaks.
In Fig.~\ref{Fig:11}  results for the West-O'Donnell power spectrum
defined by the parameters $k_-=1.61(\omega/c)$ and $k_+=2.76(\omega/c)$
(which gives $s=0.658$ and $\left<D\right>=137.3$~nm) are  presented.
In this case the satellite peaks are predicted to appear at
angles $\theta^\pm=\pm17.5^\circ$ for normal incidence ($\theta_0=0^\circ$)
and at $\theta^\pm=12.3^\circ,-22.8^\circ$ for $\theta_0=5^\circ$.
It should be noted that we only expect one pair of satellite peaks
because the incident wave can  directly excite only two of the
three possible guided waves ($k_->q_3(\omega)$). Furthermore,
with , $k_\pm>\omega/c$, single-scattering processes will not
contribute at all (see the discussion above).
From Fig.~\ref{Fig:11} we see that the satellite peaks in the
smoothed data are found at the expected positions both for
$\theta_0=0^\circ$ and $\theta_0=5^\circ$, even though their amplitudes
are small. This was the the same conclusion arrived at in
Ref.~\cite{Madrazo}.

\section{Conclusions}
\label{Sec:Conclusions}

We have presented a numerical study of the scattering of electromagnetic
waves of both p- and s-polarization from a system consisting of a
planar dielectric film deposited on a randomly rough  perfectly conducting
substrate.
The numerical calculation was performed by considering a finite length of
the randomly rough surface $L\gg\lambda$, then by solving the corresponding
reduced Rayleigh
equation for the scattering geometry by standard techniques. By averaging
the results for the scattering amplitude and its squared modulus over the
ensemble of realizations of the
rough surface, the mean incoherent reflection coefficient was obtained.
The numerical results for the mean differential reflection
coefficient shows that the scattering geometry under investigation
gives  rise to satellite peaks at well defined positions for most of
the scattering and roughness parameters considered, in accordance with theory.

From a methodological standpoint the present work demonstrates that a
purely numerical approach to the solution of a reduced Rayleigh equation by
standard numerical techniques, without the necessity of replicating a
segment of random surface of length $L$ periodically, as was done 
in~\cite{Madrazo}, is a viable, 
nonperturbative approach to the investigation of the
scattering of light from one-dimensional random surfaces whose roughness is
such that the Rayleigh hypothesis is valid.

From a physical standpoint the results presented in Fig.~\ref{Fig:4} 
strongly suggest that the scattering system studied 
in the present work is a more favorable
one for the experimental observation of satellite peaks than the one
studied in~\cite{Sanchez-Gil94,Sanchez-Gil96,Madrazo}, 
due to the increased incoherent scattering intensity
to which it gives rise.  In addition, structures of the kind studied in the
present work should be easier to fabricate than the one considered 
in~\cite{Sanchez-Gil94,Sanchez-Gil96,Madrazo}.

\acknowledgements

I.\ Simonsen would like to thank the Department of Physics and Astronomy,
University of California, Irvine,  for its kind hospitality during a
stay during which part of this work was conducted.
I.\ S. would also like to  thank the
Research  Council of Norway (Contract No. 32690/213)
and Norsk Hydro ASA for financial support.

The research of A.A.\ Maradudin was supported in part by Army Research
Office Grant DAAH 04-96-1-0187.

This work has received support from the Research  Council of Norway
(Program for Supercomputing) through a grant of computing time.


\newpage

\widetext

\begin{figure}
    \caption{The scattering geometry considered in the present
        paper. The medium above the illuminated planar surface at
        $x_3=d$ is vacuum. The randomly rough surface at
        $x_3=\zeta(x_1)$ is characterized by a stationary, zero-mean, 
        Gaussian random process $\zeta (x_1)$. 
        Below this surface the medium is a perfect
        conductor, while a dielectric film is present in the region
        $\zeta(x_1)<x_3<d$. This film is characterized by the frequency
        dependent dielectric function $\varepsilon(\omega)$.
        The angles of incidence and scattering  are $\theta_0$ and $\theta$,
        as indicated in the figure.}
    \label{Fig:1}
\end{figure}

\begin{figure}
    \caption{Two rough profiles both with a Gaussian height
        distribution with an {\sc rms}-value $\sigma=30$~nm.
        The power spectrum is of the (a) Gaussian type, with
        $a=100$~nm, and the (b) West-O'Donnell type, with
        $k_{-}=0.82(\omega/c)$ and  $k_{+}=1.97(\omega/c)$.
        Here the wavelength is  $\lambda = 633$~nm.
        With these parameters the {\sc rms}-slope and distance between
        consecutive peaks and valleys are respectively $s=0.424$,
        $\left<D\right>=128.3$~nm, and $s=0.427$,
        $\left<D\right>=201.1$~nm for the case of the Gaussian and
        West-O'Donell power spectra.
        Note that there are different scales on the first and second
        axes, with the result that the profiles appear much
        rougher than they really are.}
    \label{Fig:2}
\end{figure}

\begin{figure}
    \caption{The  contribution to the mean differential
        reflection coefficient from the incoherent component of the
scattered light  $\left< \partial R_s / \partial \theta
        \right>_{\mathrm incoh}$ as a function of the scattering angle
        $\theta$ when an s-polarized plane wave of wavelength
        $\lambda =633$~nm is  incident normally
        ($\theta_0=0^\circ$) on the scattering system
        depicted in Fig.~\protect\ref{Fig:1}.
        The dielectric function of the film is
        $\varepsilon(\omega) =2.6896+i0.01$,
        and its  mean thickness is $d=500$~nm.
        The surface profile function $\zeta(x_1)$ is characterized by
        a Gaussian surface height distribution with an {\sc rms}-height,
        $\sigma = 30$~nm, and a Gaussian  surface-height 
        autocorrelation function with a $1/e$-correlation length  
        of $a=100$~nm.
        The scattering system supports satellite peaks at
        $\pm 17.7^{\circ}$ as indicated by
        the vertical dashed lines in the figure.
        The raw data are presented in Fig.~\protect\ref{Fig:3}a, 
        while the their (five-point filter) smoothed analogs are 
        given in Fig.~\protect\ref{Fig:3}b.}
        \label{Fig:3}
\end{figure}

\begin{figure}
    \caption{The  contribution to the mean differential
        reflection coefficient from the incoherent component of the
        scattered light  $\left< \partial R_s / \partial \theta
        \right>_{\mathrm incoh}$ as a function of the scattering angle
        $\theta$ when an s-polarized plane wave of wavelength
        $\lambda =633$~nm is  incident normally
        ($\theta_0=0^\circ$) on the scattering system
        depicted in Fig.~\protect\ref{Fig:1}.
        The dielectric function of the film is
        $\varepsilon(\omega) =2.6896+i0.01$,
        and its  mean thickness is $d=500$~nm.
        The surface roughness profile function $\zeta(x_1)$ is 
        characterized by a Gaussian surface height distribution 
        with an {\sc rms}-height $\sigma = 30$~nm and a West-O'Donnell 
        power spectrum defined by the parameters $k_-=0.82 (\omega/c)$
        and $k_+=1.97 (\omega/c)$.
        The scattering system supports satellite peaks at
        $\pm 17.7^{\circ}$ as indicated by
        the vertical dashed lines in the figure.
        The raw data for for the geometry considered are presented
        in Fig.~\protect\ref{Fig:4}a.
        The numerical results obtained by Madrazo and 
        Maradudin~\protect\cite{Madrazo} for a corresponding 
        geometry with the vacuum-dielectric interface being the 
        rough one (see text) are given in Fig.~\protect\ref{Fig:4}b.}
        \label{Fig:4}
\end{figure}

\begin{figure}
    \caption{The same as Fig.~\protect\ref{Fig:4}a (raw data), but now
        using the (a) Fourier- and the (b) Taylor-methods for calculating
        $I(\gamma |q)$ (see text).
        In Fig.~\protect\ref{Fig:6}c the difference between the results for the
        contributions to the mean differential reflection coefficient 
        calculated by these two methods,
        $\Delta\left< \partial R_s / \partial \theta
        \right>_{\mathrm incoh}$, is presented. We observe that the relative
        discrepancy is roughly of the order of 1\% for the parameters used.
        The number of samples used in obtaining the results was
        $N_\zeta=50$, and the number of spatial discretization points was
         $N=2^{11}=2048$ in order to take advantage of the Fast
        Fourier transform. For the two methods the same surface
        profiles have been used in the calculation.
        The other parameters are the one used in obtaining the results
        plotted in Fig.~\protect\ref{Fig:4}.}
    \label{Fig:6}
\end{figure}

\begin{figure}
    \caption{The same as Fig.~\protect\ref{Fig:4}a (raw data),
        but now for an angle of incidence  $\theta_0=5^\circ$.
        The satellite peaks should now be present
        at scattering angles $\theta = 12.1^\circ$ and
        $\theta = -26.1^\circ$, as indicated by the vertical 
        dashed lines.}
    \label{Fig:5}
\end{figure}

\begin{figure}
    \caption{The  contribution to the mean differential
        reflection coefficient from the incoherent component of the
scattered light, $\left< \partial R_p / \partial \theta
        \right>_{\mathrm incoh}$ as a function of the scattering angle
        $\theta$ when a p-polarized plane wave of wavelength
        $\lambda =633$~nm is  incident normally
        ($\theta_0=0^\circ$) on the scattering system
        depicted in Fig.~\protect\ref{Fig:1}.
        The dielectric function of the film is
        $\varepsilon(\omega) =2.6896+i0.01$,
        and its  mean thickness is $d=500$~nm.
        The surface profile function $\zeta(x_1)$, is characterized by
        a Gaussian surface height distribution with an {\sc rms}-height,
        $\sigma = 30$~nm and a Gaussian surface-height 
        autocorrelation function with a $1/e$-correlation 
        length  of $a=100$~nm.
        The scattering system may give raise to six satellite peaks at
        $\theta_{(1,2)}^\pm=\pm 13.3^\circ$,
        $\theta_{(2,3)}^\pm=\pm 22.3^\circ$ and
        $\theta_{(1,3)}^\pm=\pm 37.6^\circ$ as indicated by
        the vertical dashed lines in the figure.
        The raw data are presented in Fig.~\protect\ref{Fig:7}a, 
        while the their (five-point filtered) smoothed analogs 
        are given in Fig.~\protect\ref{Fig:7}b.}
        \label{Fig:7}
\end{figure}

\begin{figure}
    \caption{The same as Fig.~\protect\ref{Fig:7}, but now for
        for a West-O'Donnell power spectrum defined by
        $k_-=0.82 (\omega/c)$ and $k_+=1.97 (\omega/c)$.}
    \label{Fig:8}
\end{figure}

\begin{figure}
    \caption{The same as Fig.~\protect\ref{Fig:8}a, but now for an
         angle of incidence of $\theta_0=5^\circ$.}
    \label{Fig:9}
\end{figure}

\begin{figure}
    \caption{The  contribution to the mean differential
        reflection coefficient from the incoherent component of the
        scattered light $\left< \partial R_p / \partial \theta
        \right>_{\mathrm incoh}$ as a function of the scattering angle
        $\theta$ when a p-polarized plane wave of wavelength
        $\lambda =633$~nm is normally incident
        ($\theta_0=0^\circ$) on the scattering system
        depicted in Fig.~\protect\ref{Fig:1}.
        The dielectric function of the film is
        $\varepsilon(\omega) =5.6644+i0.01$,
        and its  mean thickness is $d=380$~nm.
        The surface profile function $\zeta(x_1)$, is characterized by
        a Gaussian surface height distribution with an {\sc rms}-height,
        $\sigma = 30$~nm and a Gaussian surface-height autocorrelation 
        function with a $1/e$-correlation length  of $a=100$~nm.
        The scattering system supports six satellite peaks (see
        Ref.~\protect\cite{Madrazo}), but two of them are in the
        non-radiative part of the spectrum. The real peaks should
        appear at $\theta^\pm_{(1,2)}=\pm17.5^\circ$ and
        $\theta^\pm_{(2,3)}=\pm46.1^\circ$.
        Notice that only the smoothed data are given.}
        \label{Fig:10}
\end{figure}

\begin{figure}
    \caption{The same as Fig.~\protect\ref{Fig:10}, but now
        for a West-O'Donnell power spectrum defined by
        $k_-=1.61 (\omega/c)$ and $k_+=2.76 (\omega/c)$ for an angle of
        incidence $\theta_0=0^\circ$ (a) and $\theta_0=5^\circ$ (b).
        The real satellite peaks are predicted to appear at
        $\theta^\pm=\pm 17.5^\circ$ ($\theta_0=0^\circ$) and
        $\theta^\pm=12.3^\circ,-22.8^\circ $ ($\theta_0=5^\circ$).
        Only the smoothed data are presented.}
    \label{Fig:11}
\end{figure}

\newpage
\setcounter{figure}{1}
\newcommand{\mycaption}[2]{\begin{center}{\bf Figure
            \thefigure}\\{#1}\\{\em
            #2}\end{center}\addtocounter{figure}{1}}
\newcommand{\myauthor}{I.\ Simonsen and A.\ A.\ Maradudin}
\newcommand{\mytitle}{Numerical simulation of electromagnetic 
    wave scattering from planar dielectric films deposited 
    on rough perfectly conducting substrates}

\begin{figure}
    \begin{center}
        \begin{tabular}{@{}c@{\hspace{1.0cm}}c@{}}
            \epsfig{file=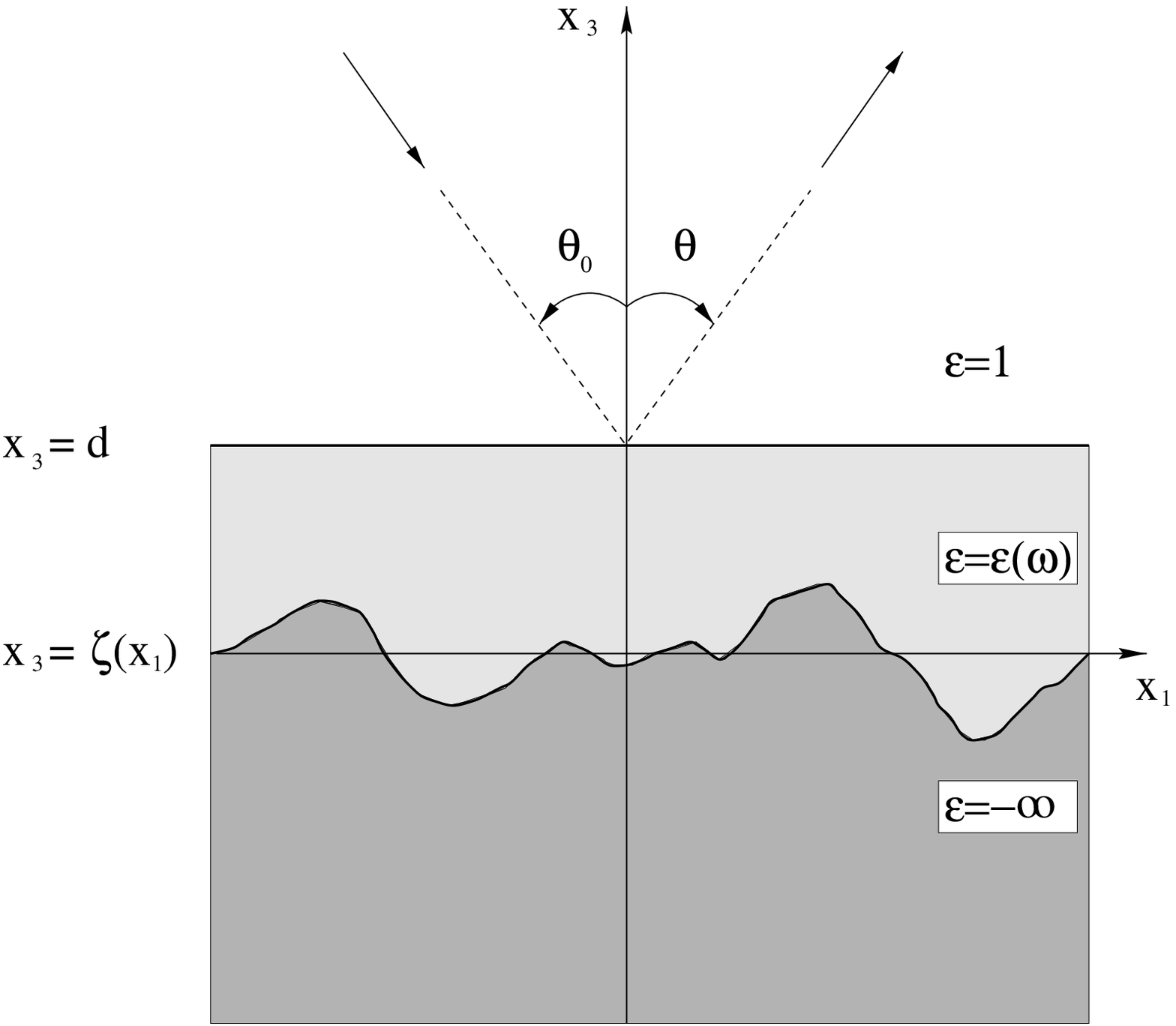,
                       width=8.5cm,height=8.5cm} &
        \end{tabular}
    \end{center}
    \mycaption{\myauthor}{\mytitle}
\end{figure}

\begin{figure}
    \begin{center}
        \begin{tabular}{@{}c@{\hspace{1.0cm}}c@{}}
            \epsfig{file=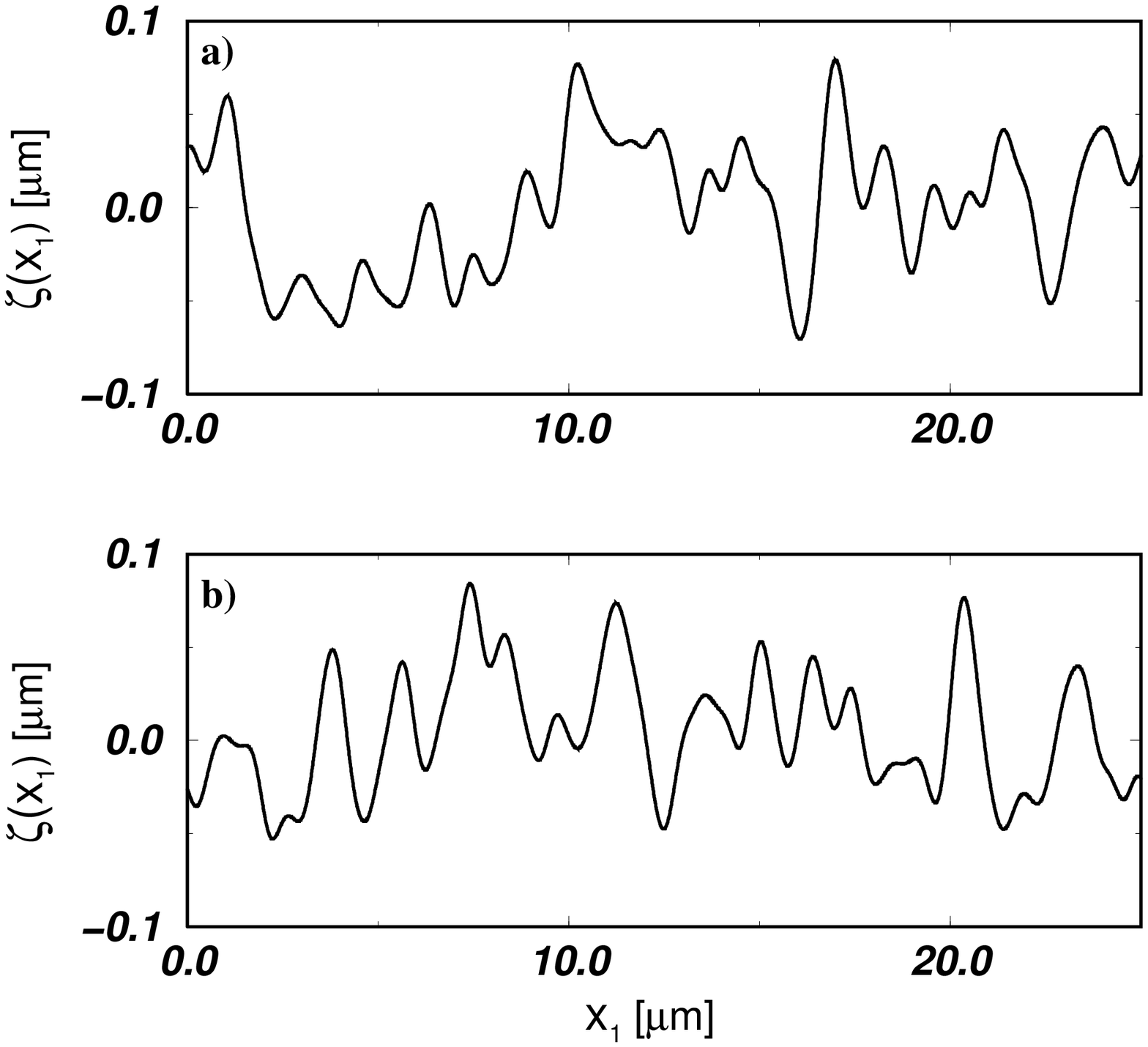,
                       width=8.5cm,height=8.5cm} &
        \end{tabular}
    \end{center}
    \mycaption{\myauthor}{\mytitle}
\end{figure}

\begin{figure}
    \begin{center}
        \begin{tabular}{@{}c@{\hspace{1.0cm}}c@{}}
            \epsfig{file=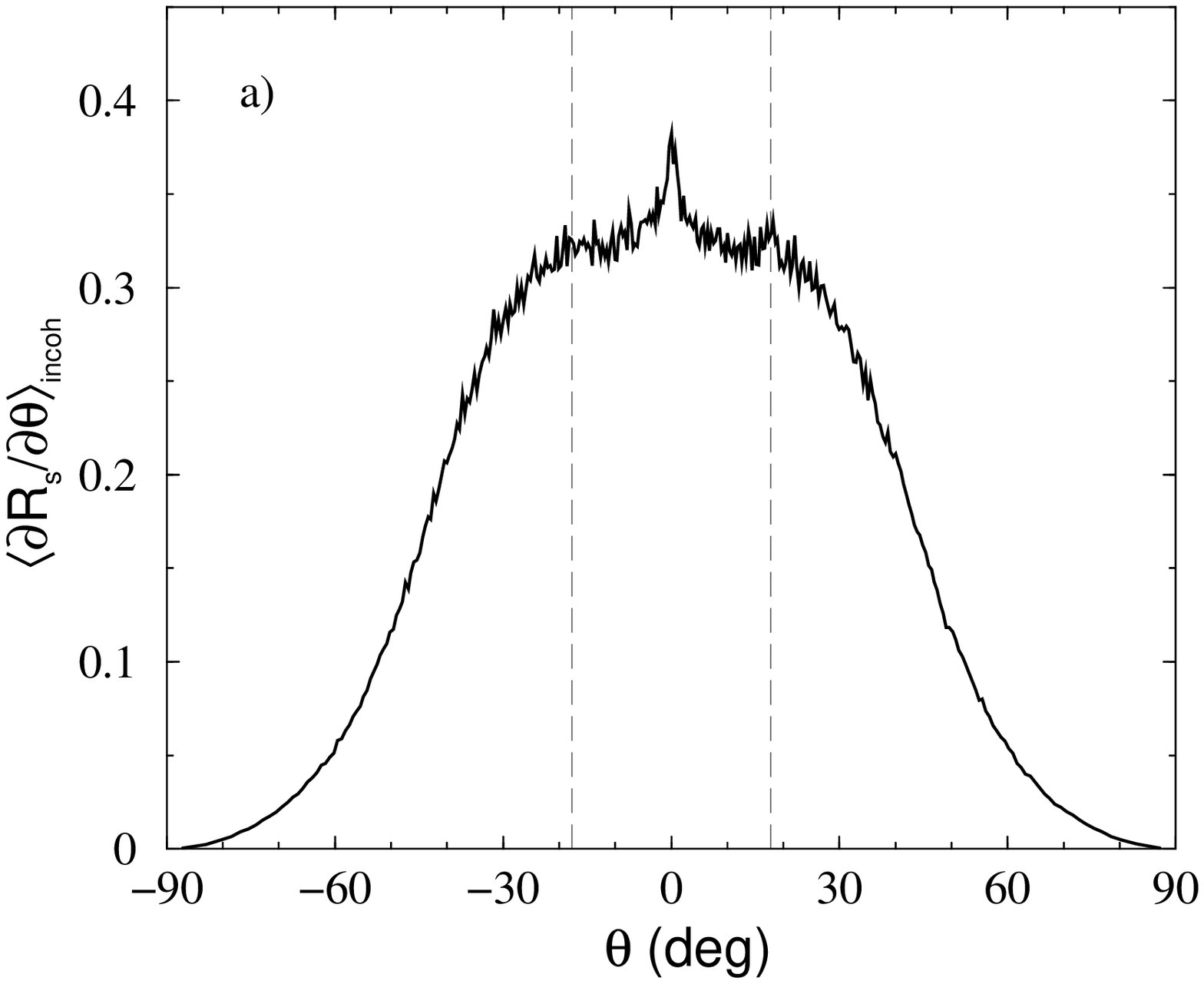,
                       width=8.5cm,height=8.5cm} &
            \epsfig{file=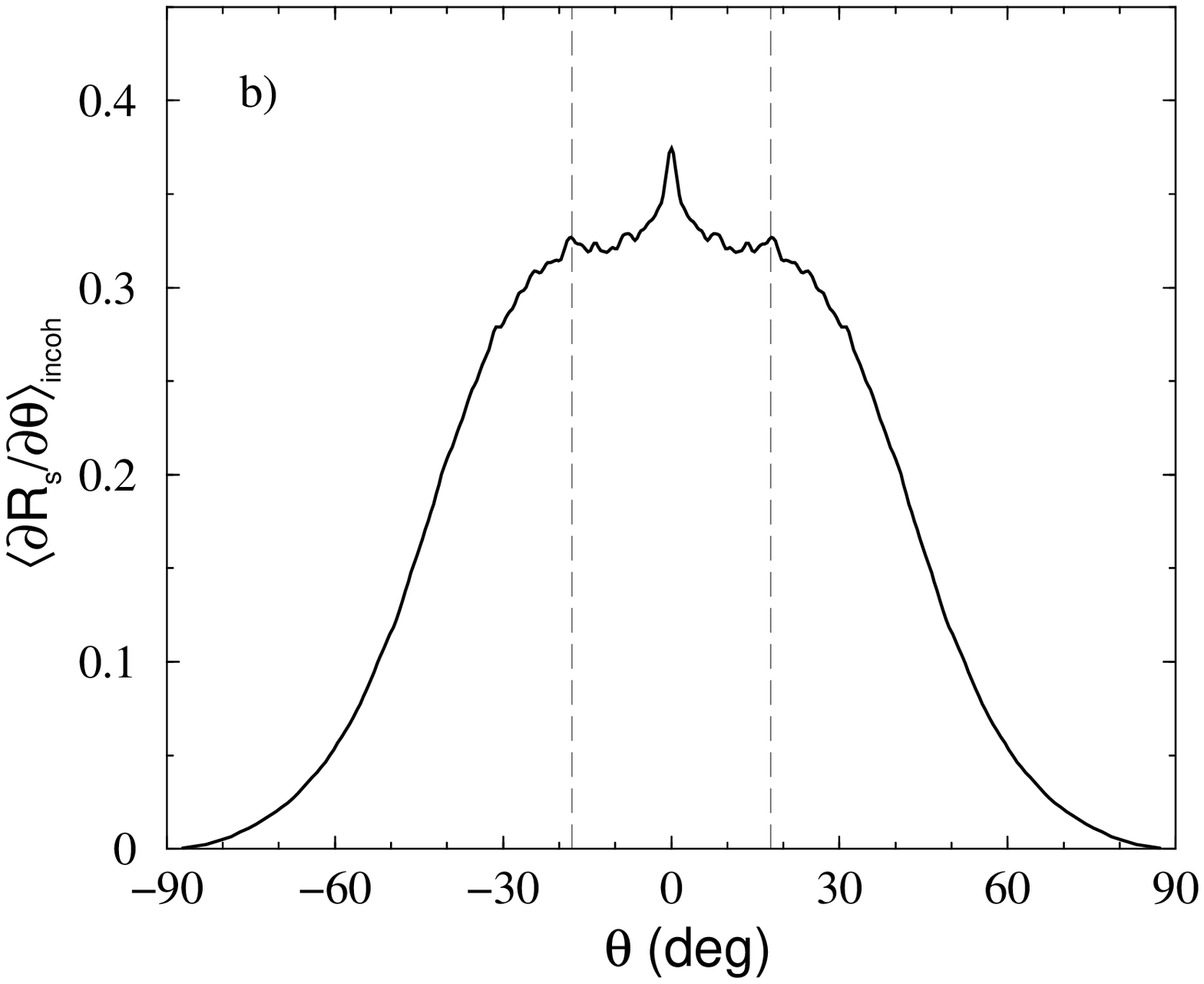,
                       width=8.5cm,height=8.5cm}
        \end{tabular}
    \end{center}
    \mycaption{\myauthor}{\mytitle}
\end{figure}

\begin{figure}
    \begin{center}
        \begin{tabular}{@{}c@{\hspace{1.0cm}}c@{}}
            \epsfig{file=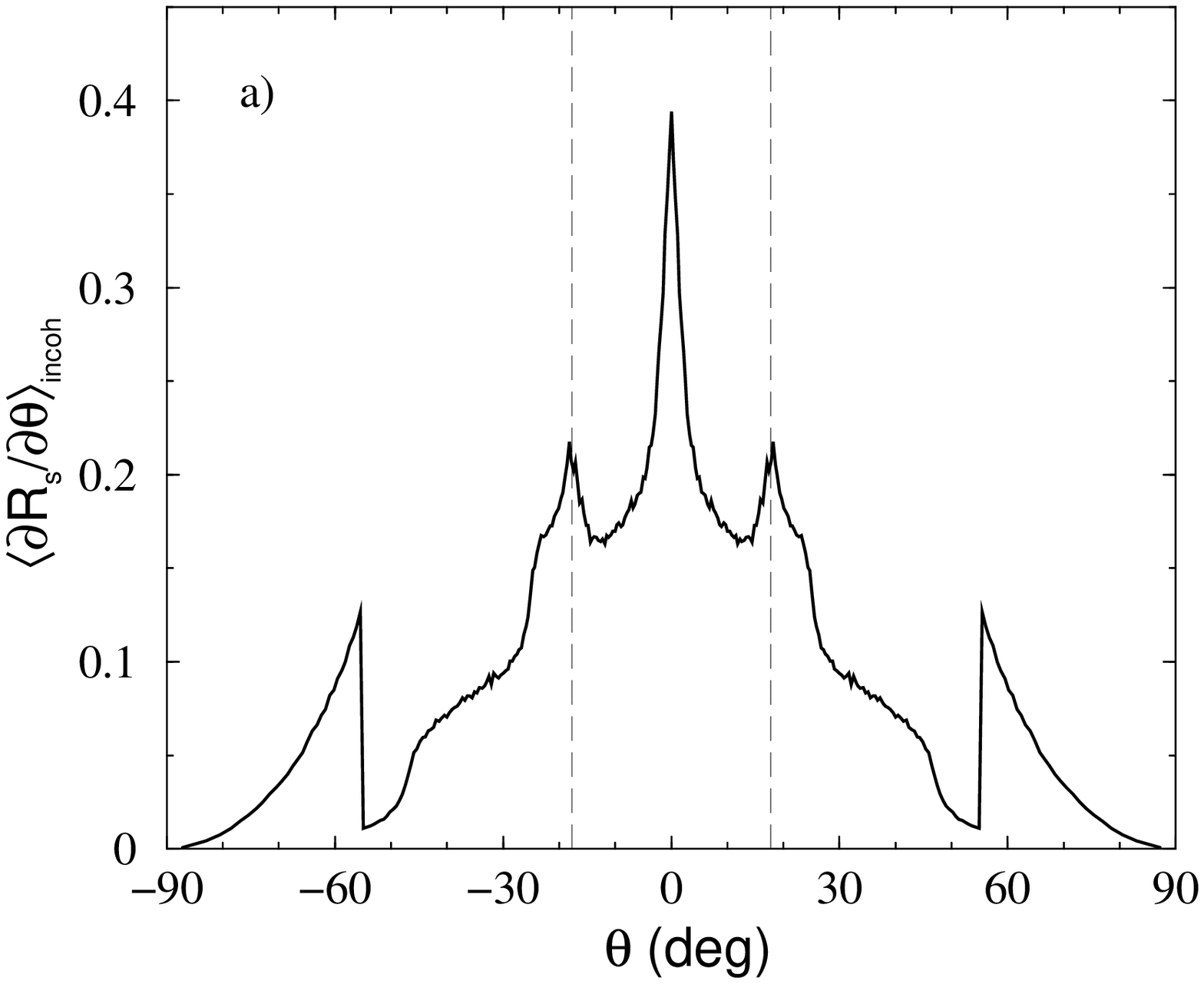,
                       width=8.5cm,height=8.5cm} &
            \epsfig{file=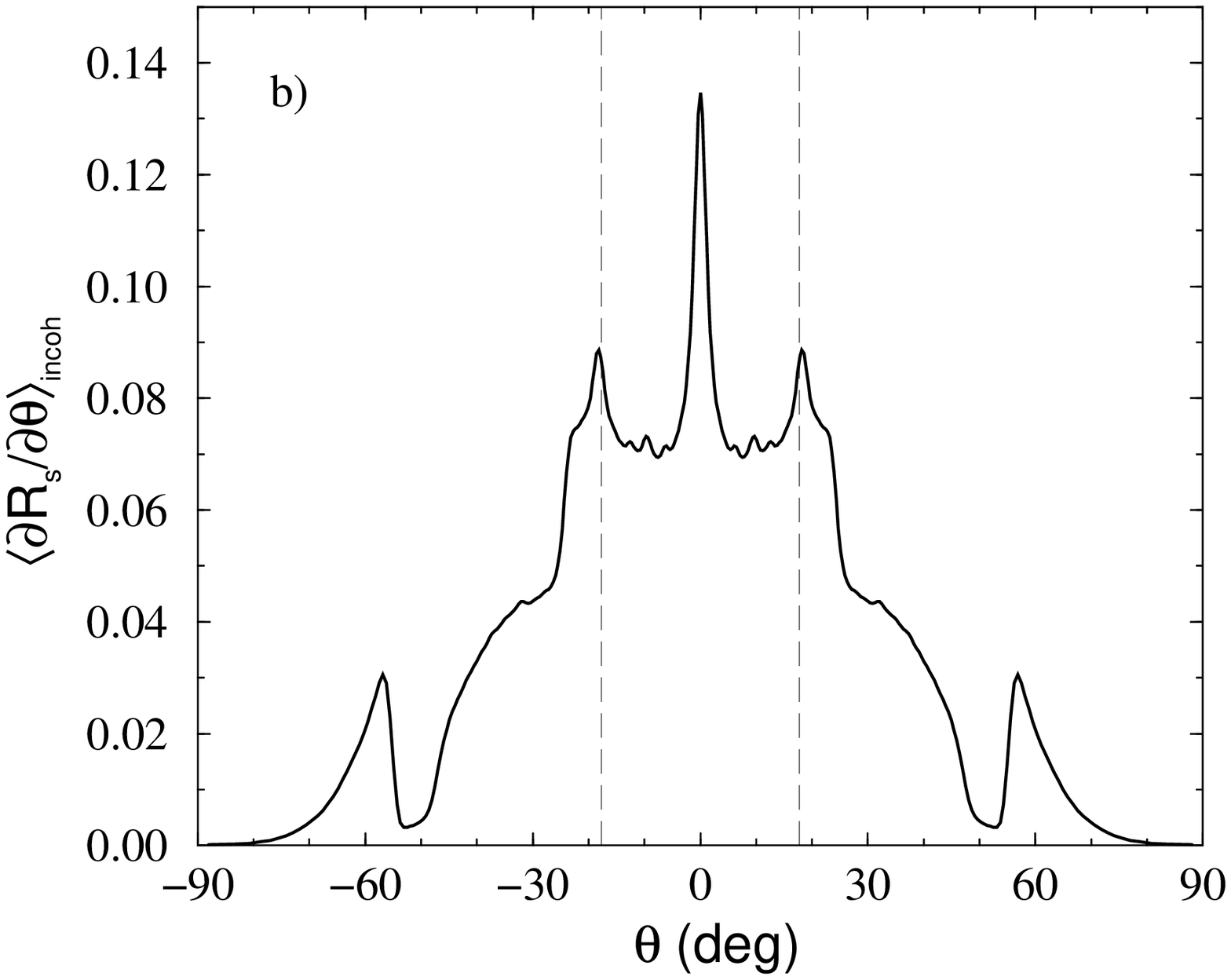,
                       width=8.5cm,height=8.5cm}
        \end{tabular}
    \end{center}
    \mycaption{\myauthor}{\mytitle}
\end{figure}

\begin{figure}
    \begin{center}
        \begin{tabular}{@{}c@{\hspace{1.0cm}}c@{}}
            \epsfig{file=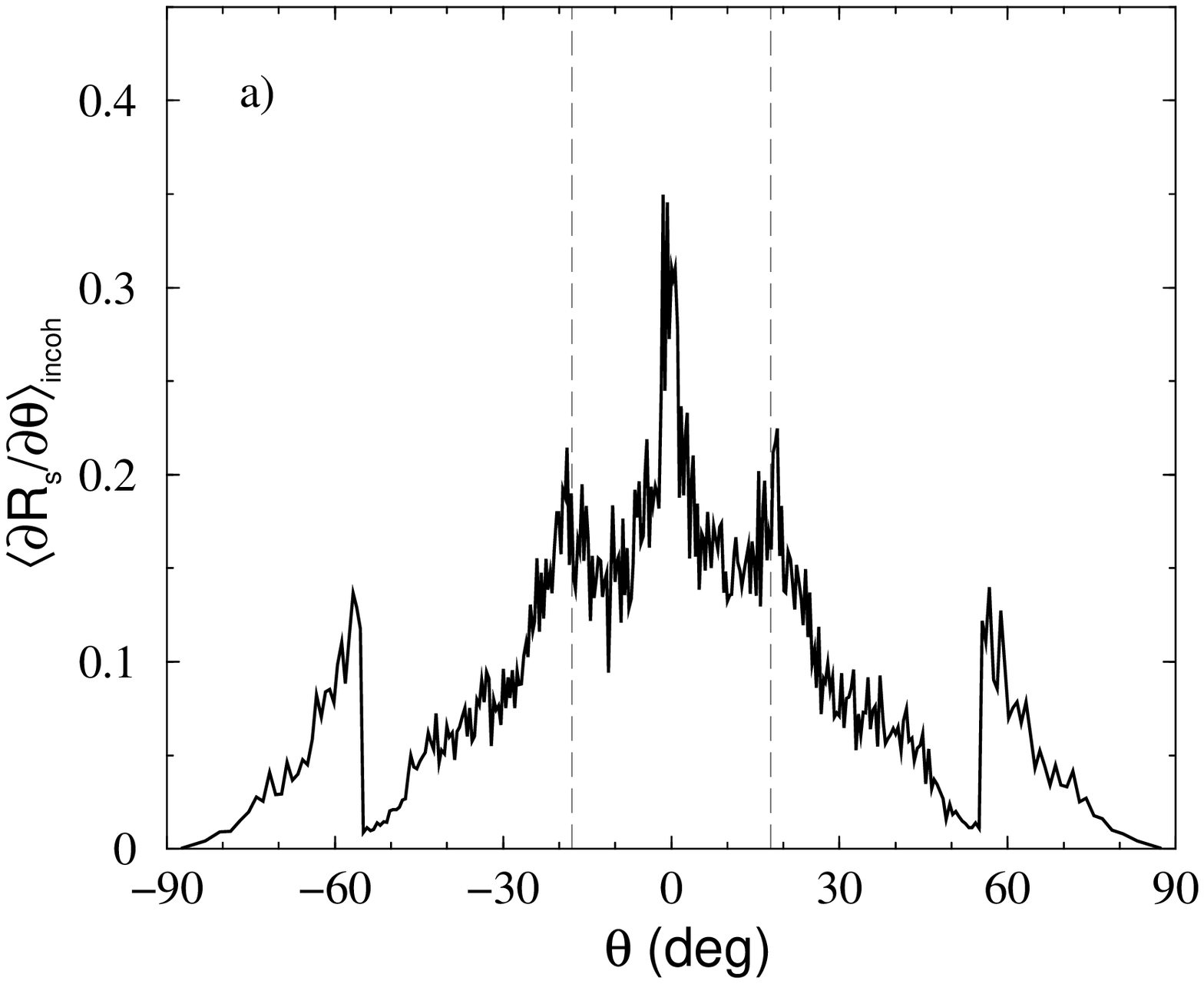,
                       width=8.5cm,height=8.5cm} &
            \epsfig{file=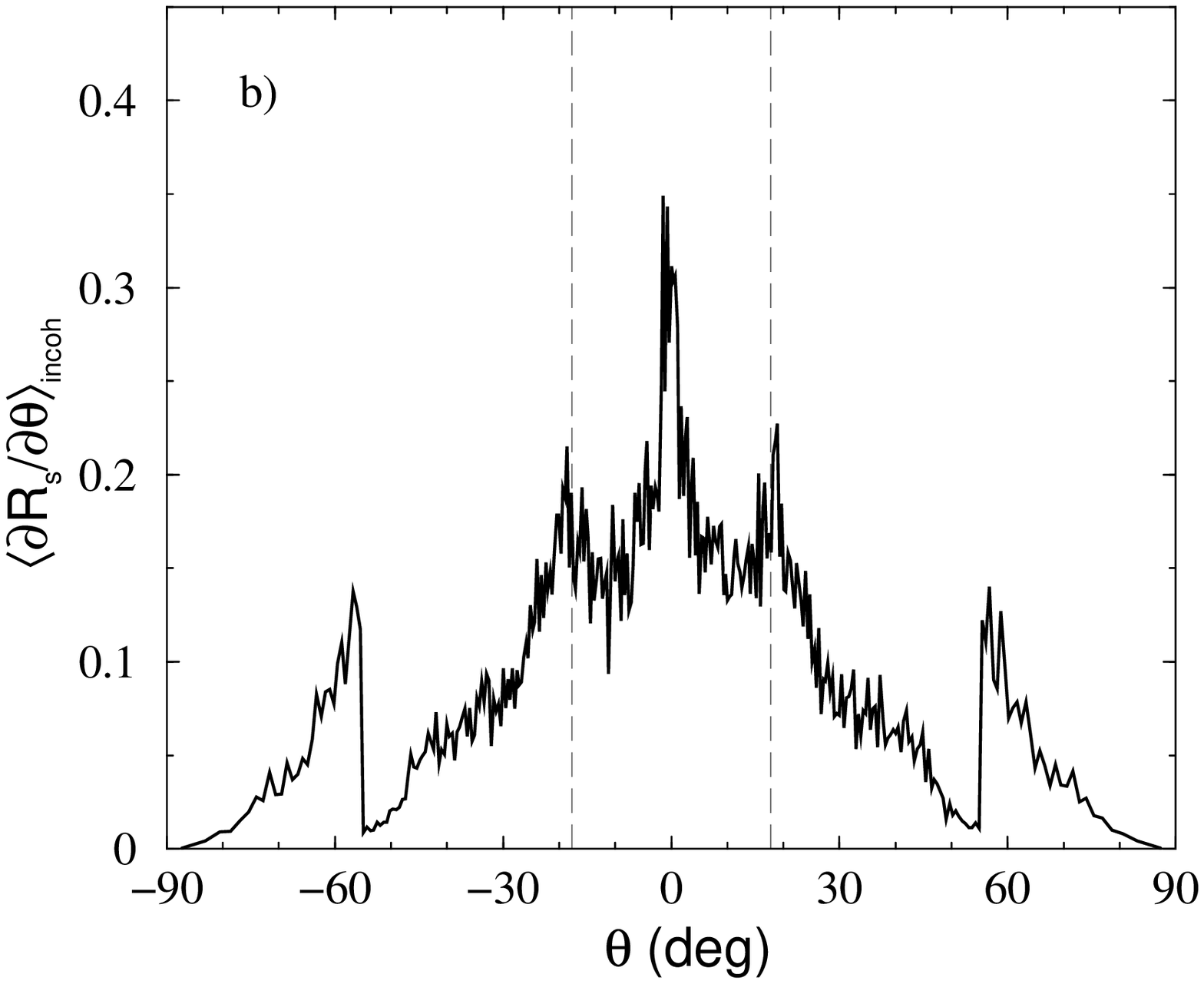,
                       width=8.5cm,height=8.5cm} \\
            \epsfig{file=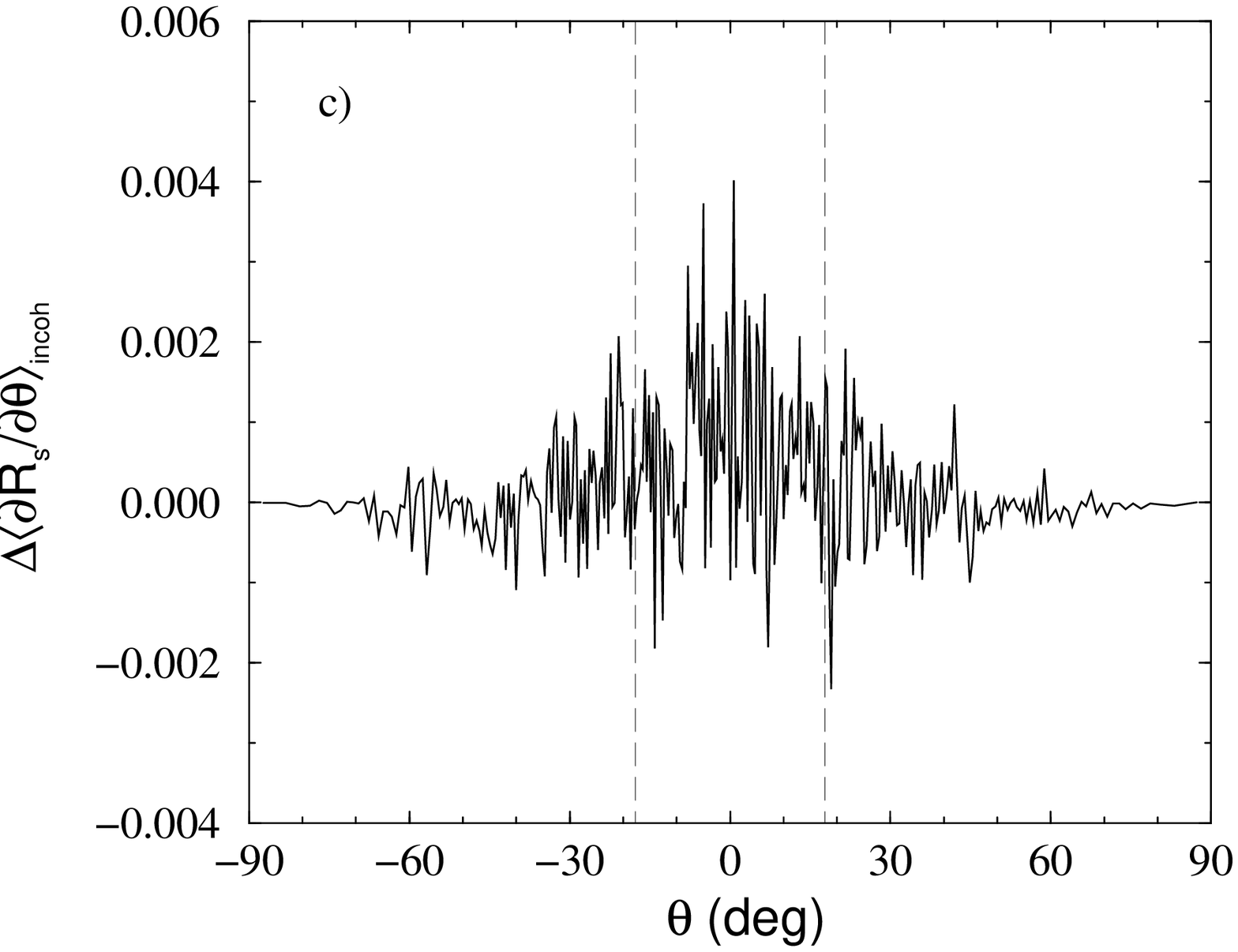,
                       width=8.5cm,height=8.5cm} &
        \end{tabular}
    \end{center}
    \mycaption{\myauthor}{\mytitle}
\end{figure}

\begin{figure}
    \begin{center}
        \begin{tabular}{@{}c@{\hspace{1.0cm}}c@{}}
            \epsfig{file=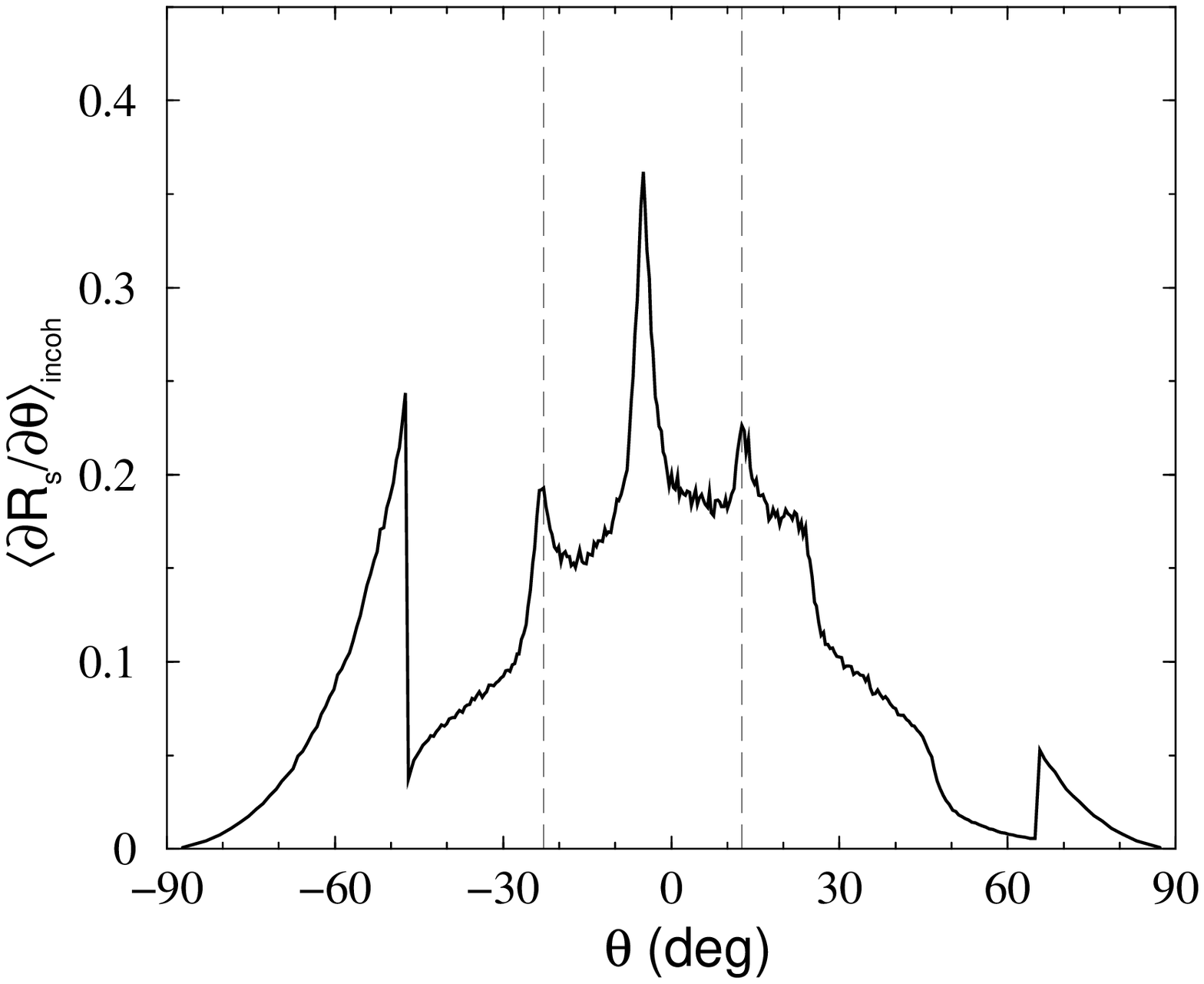,
                       width=8.5cm,height=8.5cm} &
        \end{tabular}
    \end{center}
    \mycaption{\myauthor}{\mytitle}
\end{figure}

\begin{figure}
    \begin{center}
        \begin{tabular}{@{}c@{\hspace{1.0cm}}c@{}}
            \epsfig{file=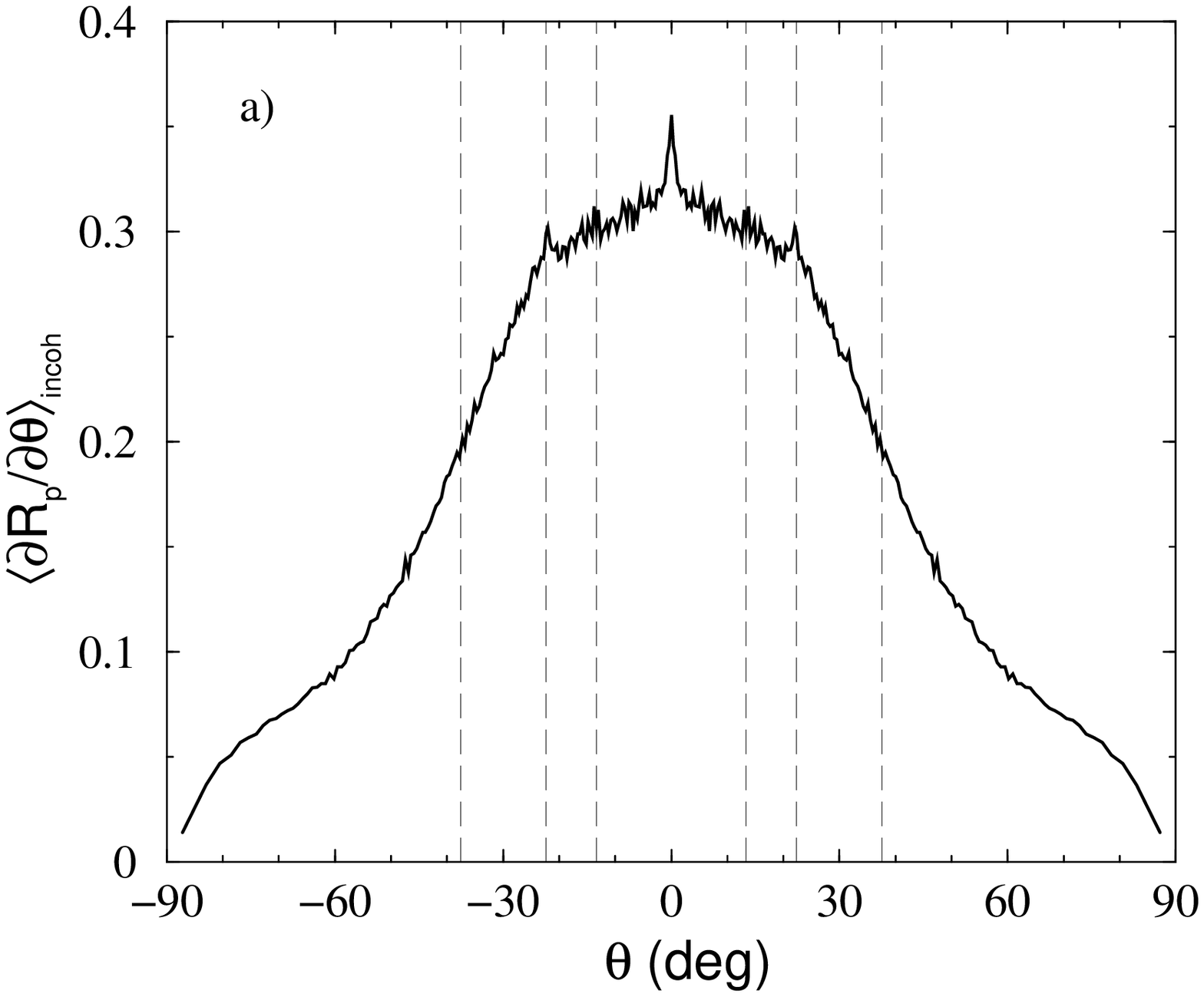,
                       width=8.5cm,height=8.5cm} &
            \epsfig{file=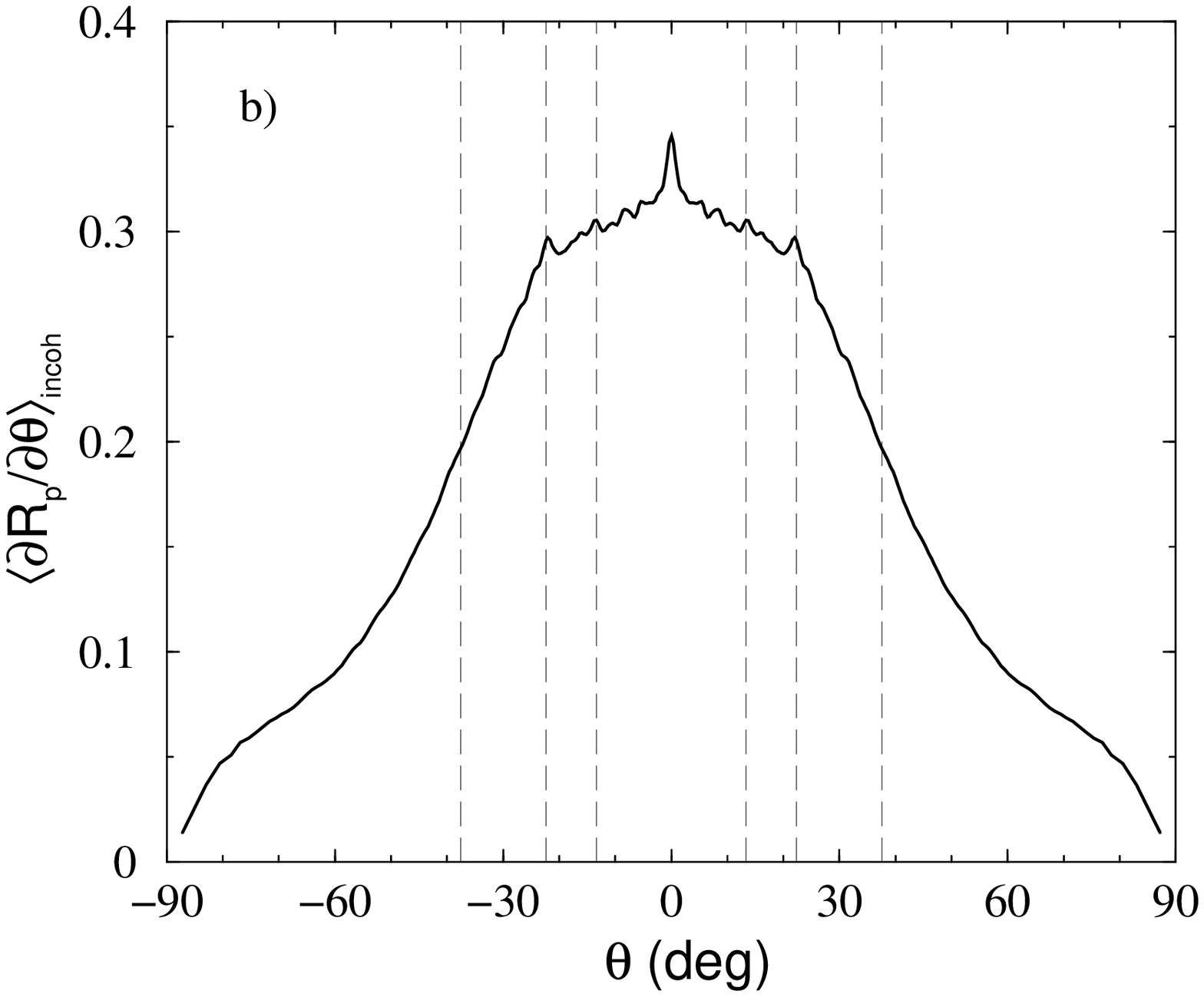,
                       width=8.5cm,height=8.5cm}
        \end{tabular}
    \end{center}
    \mycaption{\myauthor}{\mytitle}
\end{figure}

\begin{figure}
    \begin{center}
        \begin{tabular}{@{}c@{\hspace{1.0cm}}c@{}}
            \epsfig{file=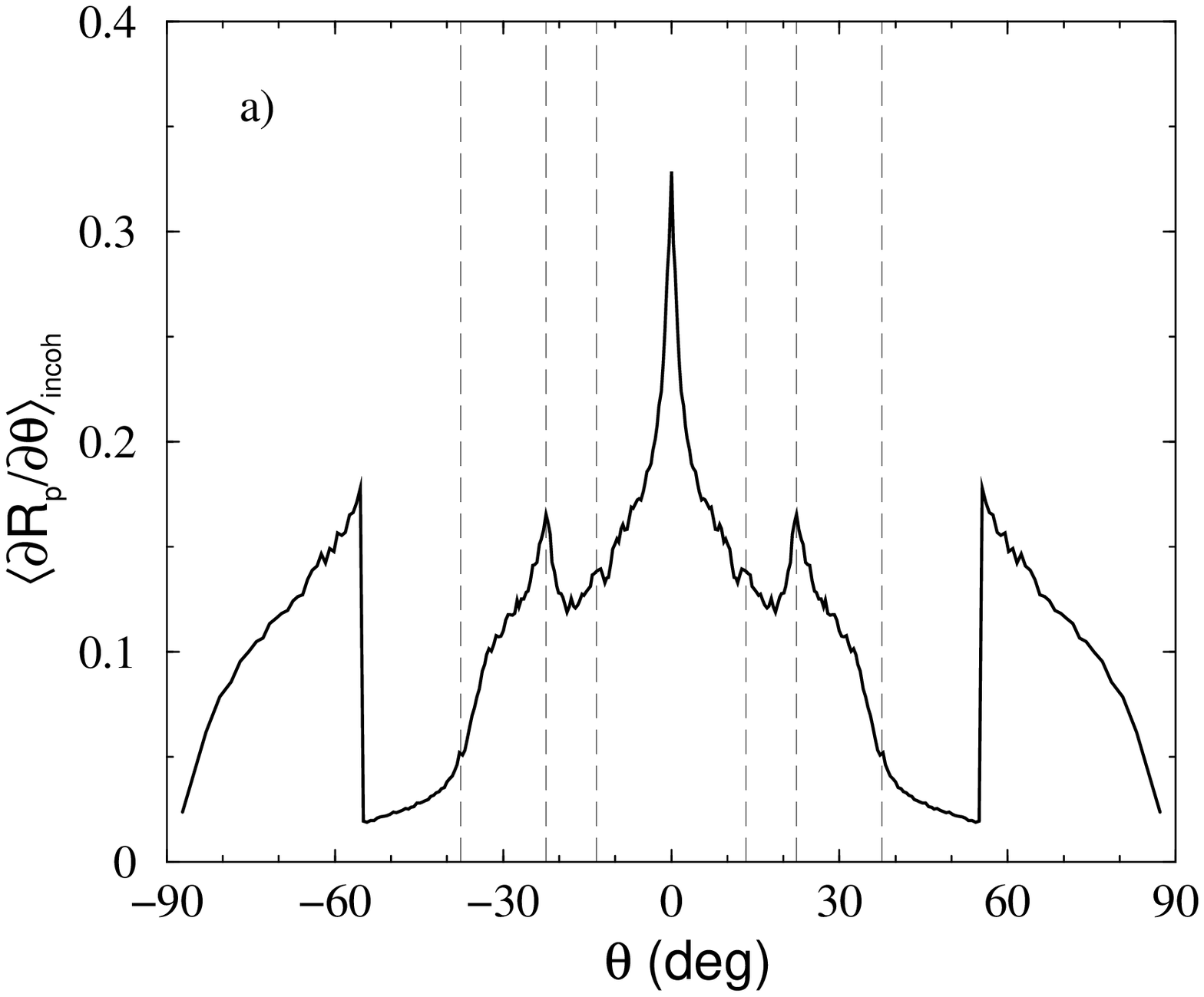,
                       width=8.5cm,height=8.5cm} &
            \epsfig{file=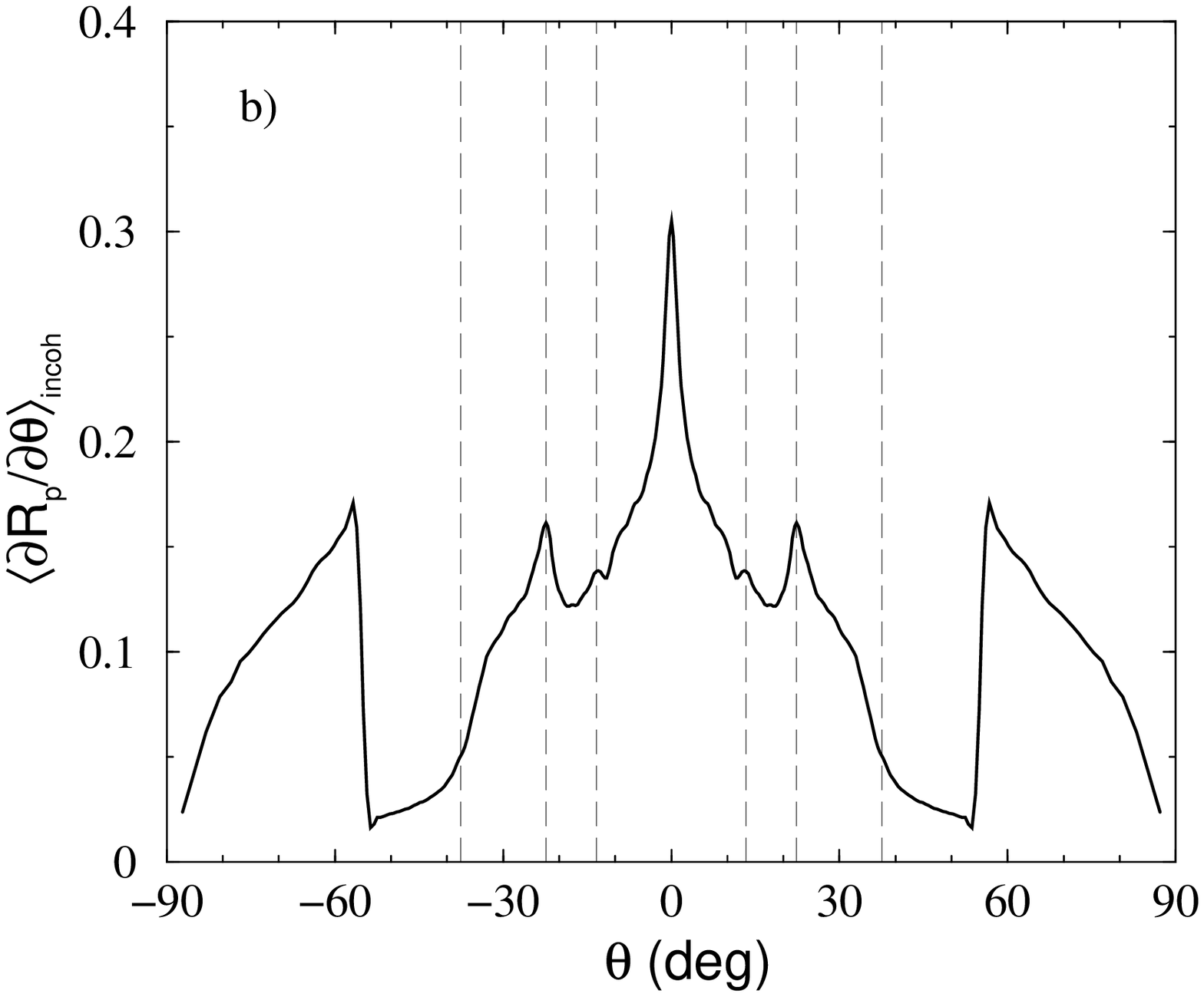,
                       width=8.5cm,height=8.5cm}
        \end{tabular}
    \end{center}
    \mycaption{\myauthor}{\mytitle}
\end{figure}

\begin{figure}
    \begin{center}
        \begin{tabular}{@{}c@{\hspace{1.0cm}}c@{}}
            \epsfig{file=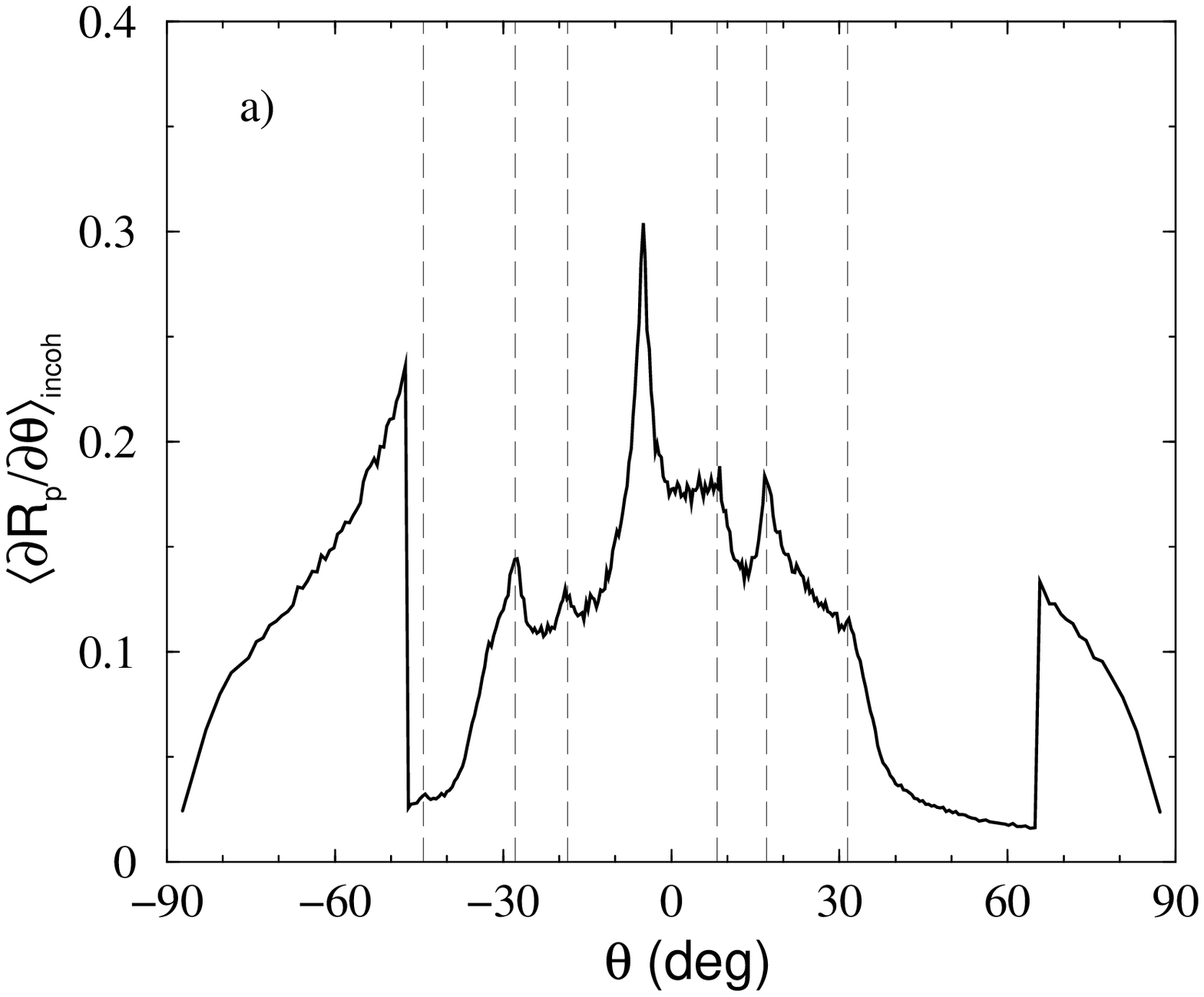,
                       width=8.5cm,height=8.5cm} &
            \epsfig{file=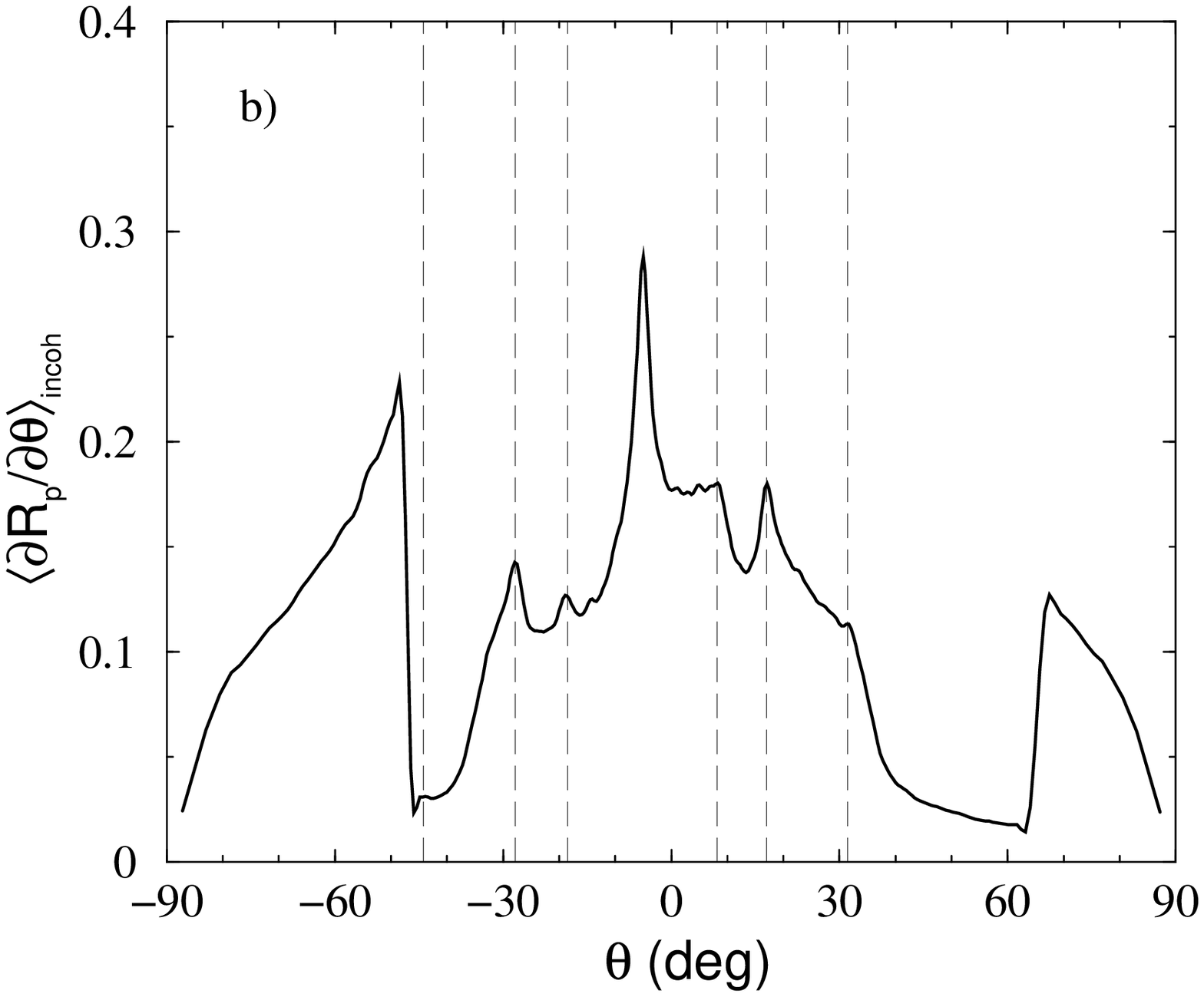,
                       width=8.5cm,height=8.5cm}
        \end{tabular}
    \end{center}
    \mycaption{\myauthor}{\mytitle}
\end{figure}

\begin{figure}
    \begin{center}
        \begin{tabular}{@{}c@{\hspace{1.0cm}}c@{}}
            \epsfig{file=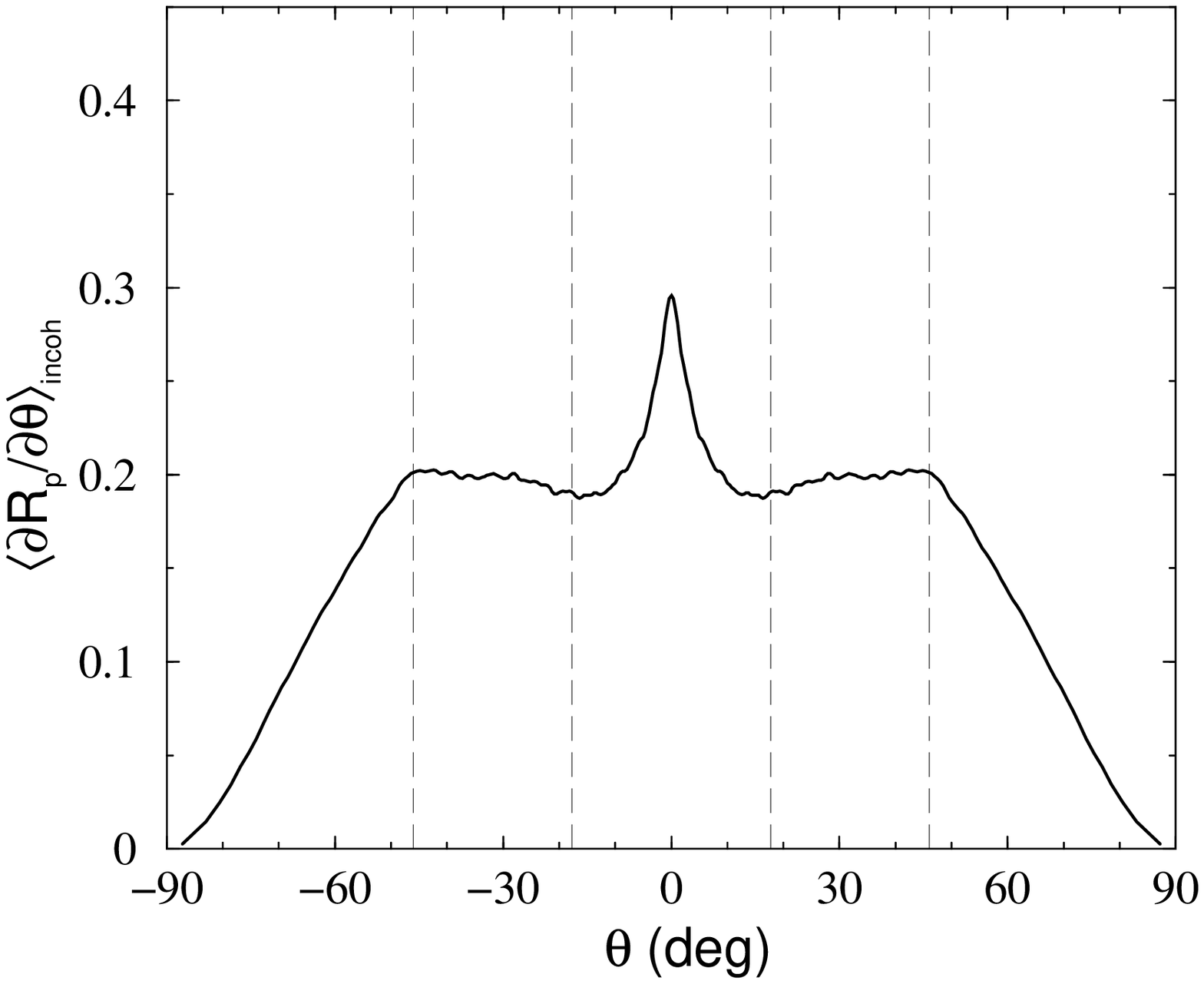,
                       width=8.5cm,height=8.5cm} &
        \end{tabular}
    \end{center}
    \mycaption{\myauthor}{\mytitle}
\end{figure}

\begin{figure}
    \begin{center}
        \begin{tabular}{@{}c@{\hspace{1.0cm}}c@{}}
            \epsfig{file=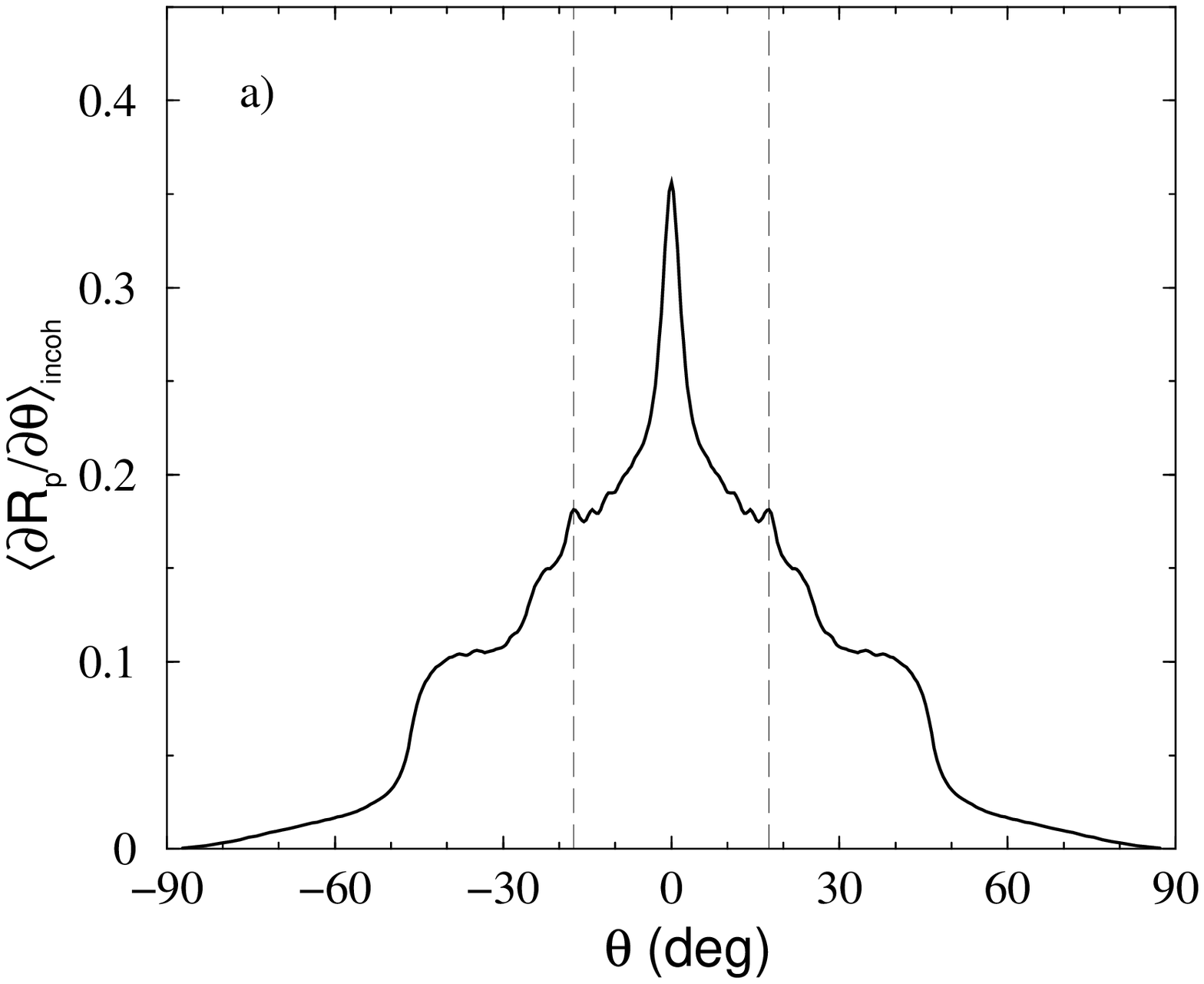,
                       width=8.5cm,height=8.5cm} &
            \epsfig{file=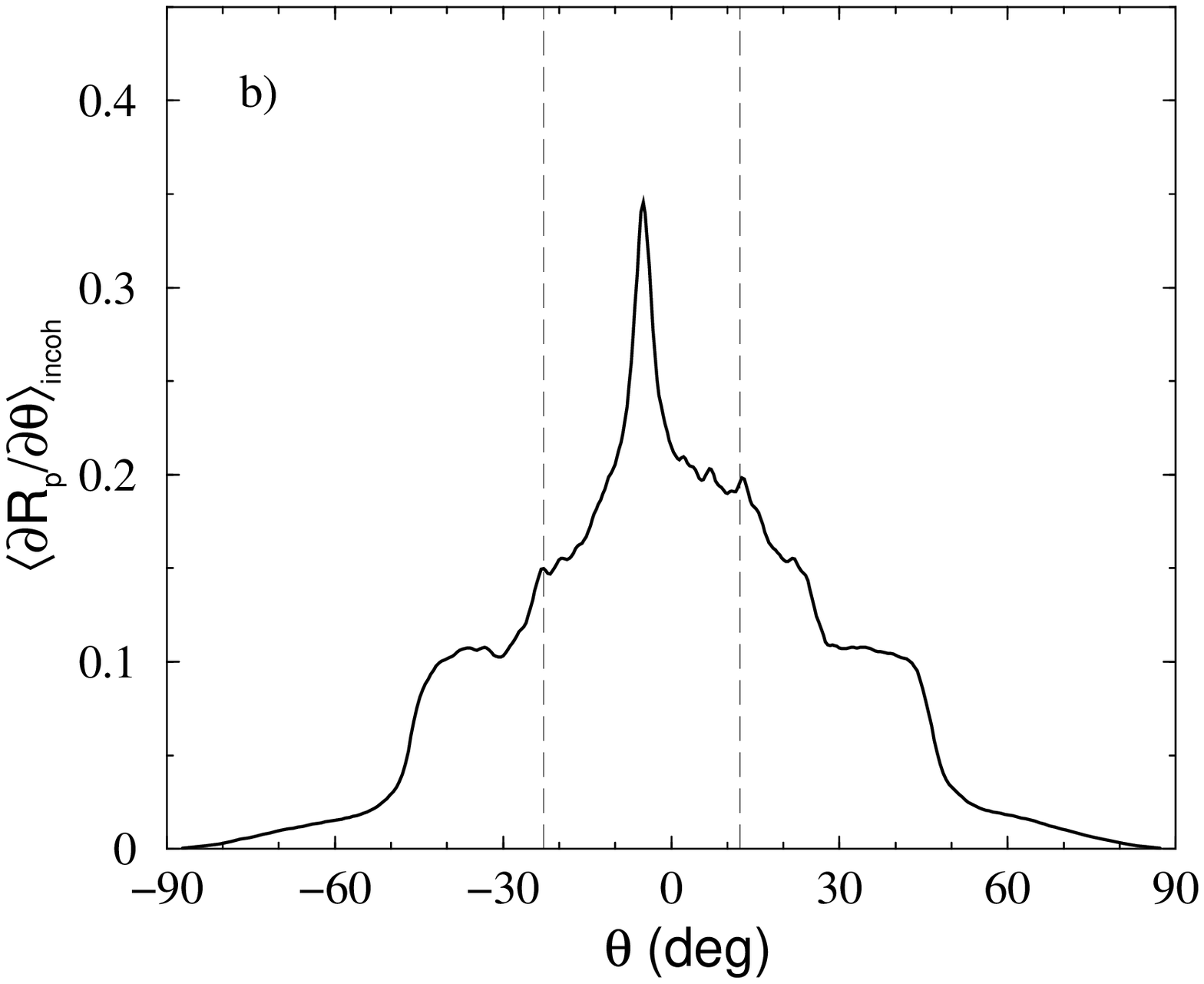,
                       width=8.5cm,height=8.5cm}
        \end{tabular}
    \end{center}
    \mycaption{\myauthor}{\mytitle}
\end{figure}


\begin{thebibliography}{99}
\bibitem{Jun_Lu}
    Jun Q.\ Lu, A.\ A.\ Maradudin, and T.\ Michel,
    J. Opt. Soc. Am. {\bf B8}, 311 (1991).

\bibitem{PhysRep}
    V.\ Freilikher, E.\ Kanzieper, and A.\ A.\ Maradudin,
    Phys. Rep. {\bf 288}, 127 (1997).

\bibitem{Freilikher94A}
    V.\ Freilikher, M.\ Pustilnik, and I.\ Yurkevich,
    Phys. Lett. {\bf A193}, 467 (1994).

\bibitem{Sanchez-Gil94}
    J.\ A.\ S\'anchez-Gil, A.\ A.\ Maradudin, Jun Q.\ Lu,
    V.\ D.\ Freilikher, M.\ Pustilnik, and  I.\ Yurkevich,
    Phys. Rev. {\bf B50}, 15353 (1994).

\bibitem{Wang}
    Z.\ L.\ Wang, H.\ Ogura, and N.\ Takahashi,
    Phys. Rev. {\bf B52}, 6027 (1995).

\bibitem{Sanchez-Gil96}
     J.\ A.\ S\'anchez-Gil, A.\ A.\ Maradudin, J.\ Q.\ Lu, V.\
     Freilikher, M.\ Pustilnik, and I.\ Yurkevich,
     J. Mod. Optics  {\bf 43}, 434 (1996).

\bibitem{Sanchez-Gil95}
    J.\ A.\ S\'anchez-Gil, A.\ A.\ Maradudin, J.\ Q.\ Lu, and V.\ D.\
    Freilikher,
    Phys. Rev. {\bf B51}, 17100 (1995).

\bibitem{McGurn89}
    A.\ R.\ McGurn, and A.\ A.\ Maradudin,
    Optics Comm. {\bf 72}, 279 (1989).

\bibitem{Ogura}
    H.\ Ogura and Z.\ L.\ Wang,
    Phys. Rev. {\bf B53}, 10 358 (1996).


\bibitem{Freilikher94B}
    V.\ Freilikher, M.\ Pustilnik, I.\ Yurkevich, and A.\ A.\
    Maradudin,
    Opt. Comm. {\bf 110}, 263 (1994).

\bibitem{Madrazo}
    A.\ Madrazo and A.\ A.\ Maradudin,
    Optics Comm. {\bf 137}, 251 (1997).

\bibitem{West}
    C.\ S.\ West and K.\ A.\ O'Donnell,
    J. Opt. Soc. Am. {\bf A12}, 390 (1995);
    Opt. Lett. {\bf 21}, 1 (1996).

\bibitem{Michel90}
    A.\ A.\ Maradudin and T.\ Michel,
    J.\ Stat.\ Phys.\ {\bf 58}, 485 (1990).

\bibitem{Rayleigh96}
    Lord Rayleigh, {\sl The Theory of Sound,\/} 2nd Ed. (Macmillan, London,
    1896), Vol II, pp. 89 and 297--311.

\bibitem{Rayleigh07}
    Lord Rayleigh, Proc. Roy. Soc. {\bf A79}, 399 (1907).

\bibitem{Ogilvy}
    J.\ A.\ Ogilvy, {\sl Theory of Wave Scattering from Random Rough
    Surfaces} (Hilger, Bristol, 1991), p. 39.

\bibitem{Simonsen98}
    I.\ Simonsen and A.\ A.\ Maradudin,
    in preparation.

\bibitem{Press}
    W.\ H.\ Press, S.\ A.\ Teukolsky, W.\ T.\ Vetterling, and B.\ P.\ Flannery,
    {\sl Numerical Recipes,\/} 2nd Ed. (Cambridge University Press,
    Cambridge, 1992).

\bibitem{AnnPhys}
   A.\ A.\  Maradudin,  T.\ Michel, A. R. McGurn, and E. R. M\'endez,
   Ann. Phys. (N.Y.) {\bf 203}, 255 (1990).


\bibitem{Madrazo24}
    See, for example, A.\ A.\ Maradudin, A.\ R.\ McGurn, and
    E.\ R.\ M\'endez,
    J. Opt. Soc. Am. A{\bf 12}, 2500 (1995).

\end{thebibliography}
\end{document}